%% file: ms.tex
\documentclass[numberedappendix]{emulateapj}

\usepackage{graphicx}
\usepackage{dcolumn}
\usepackage{bm}
\usepackage{epsf}
\usepackage{verbatim}
\usepackage{amsmath}
\usepackage{amsfonts}
\usepackage{ifthen}

\newboolean{incApp}
\setboolean{incApp}{false}
\newcommand{\mycond}[2]{\ifthenelse{#1}{#2}{}}
\newcommand{\myNcond}[2]{\ifthenelse{#1}{}{#2}}

\def\myfig#1{./#1}

\input{Definitions.tex}
\newcommand{\mycw}{{C_{\mbox{\scriptsize{w}}}}}

\newcommand{\myT}{\mathcal{T}}

\newcommand{\myv}{{\mbox{v}}}
\newcommand{\myw}{{\mbox{w}}}
\newcommand{\myU}{{\mbox{U}}}
\newcommand{\mypP}{{\lambda}}

\newcommand{\mypT}{{\lambda_T}}

\newcommand{\myprho}{{\lambda_\rho}}

\newcommand{\mypY}{{\alpha_Y}}

\newcommand{\mypv}{{\lambda_{\mbox{\scriptsize{v}}}}}
\newcommand{\mypw}{{\lambda_{\mbox{\scriptsize{w}}}}}

\newcommand{\mypz}{{\tilde{\lambda}_z}}

\newcommand{\mypgamma}{{\lambda_\gamma}}

\newcommand{\mymin}{{\mbox{\scriptsize min}}}
\newcommand{\mymax}{{\mbox{\scriptsize max}}}

\shorttitle{Spiral flows in cool cores}
\shortauthors{U. Keshet}

\begin{document}

\title{Spiral flows in cool-core galaxy clusters}

\author{Uri Keshet\altaffilmark{}}

\affiliation{Physics Department, Ben-Gurion University of the Negev, Be'er-Sheva 84105, Israel; ukeshet@bgu.ac.il}

\date{\today}

\begin{abstract}
We argue that bulk spiral flows are ubiquitous in the cool cores (CCs) of clusters and groups of galaxies.
Such flows are gauged by spiral features in the thermal and chemical properties of the intracluster medium, by the multi-phase properties of CCs, and by X-ray edges known as cold fronts.
We analytically show that observations of piecewise-spiral fronts impose strong constraints on the CC, implying the presence of a cold, fast flow, which propagates below a hot, slow inflow, separated by a slowly rotating, trailing, quasi spiral, tangential discontinuity surface.
This leads to the nearly logarithmic spiral pattern, two-phase plasma, $\rho\sim r^{-1}$ density (or $T\sim r^{0.4}$ temperature) radial profile, and $\sim 100\kpc$ size, characteristic of CCs.
By advecting heat and mixing the gas, such flows can eliminate the cooling problem, provided that a feedback mechanism regulates the flow.
In particular, we present a quasi-steady-state model for an accretion-quenched, composite flow, in which the fast phase is an outflow, regulated by active galactic nucleus bubbles, reproducing the observed low star formation rates and explaining some features of bubbles such as their $R_b\propto r$ size.
The simplest two-component model reproduces several key properties of CCs, so we propose that all such cores harbor a spiral flow.
Our results can be tested directly in the next few years, for example by ASTRO-H.
\end{abstract}


\maketitle

\section{Introduction}
\label{sec:Intro}

Most galaxy clusters show a central dense, cool core (CC) in which the radiative cooling time of the intracluster medium (ICM) drops well below the age of the cluster. Some steady, smoothly distributed heat mechanism is needed in order to balance the observed cooling, and to sustain the mild (factor of a few, typically) temperature drop towards the center. For reviews of such CC clusters (CCCs), see \cite{PetersonFabian06, McNamaraNulsen07, Soker10}.

Energetically, core cooling can plausibly be suppressed, for example by thermal conduction \citep{ZakamskaNarayan03} modified by heat buoyancy instabilities \citep{Quataert08} or by the energy output of the active galactic nucleus (AGN) in the central cD galaxy \citep[\eg][]{BirzanEtAl04}.
In particular, sufficient mechanical energy is thought to be deposited in hot AGN bubbles \citep{ChurazovEtAl02}, found in at least $70\%$ of the cool cores \citep{DunnFabian06}.
However, it is unclear if the energy can be transferred steadily and homogeneously throughout the cooling plasma, in order to stem the local thermal instability.
Moreover, a feedback mechanism is needed in order to regulate the heating and adapt to changes in the core.
If the AGN output is regulated by the accreted, cooling plasma, it could furnish such a feedback loop and quench the global thermal instability as well \citep[\eg][]{RaffertyEtAl08}.

In recent years, high resolution X-ray observations have revealed the presence of spiral patterns in a significant fraction of the CCCs for which high quality data are available \citep{ClarkeEtAl04, TanakaEtAl06, SandersEtAl09, JohnsonEtAl10, LaganaEtAl10, RandallEtAl10, BlantonEtAl11, RoedigerEtAl11}. These observations suggest a spiral morphology composed of spatially alternating, low entropy and high entropy plasma phases, which are approximately at pressure equilibrium.
The low entropy component is colder and denser (both by a similar factor $q$ of up to a few), and higher in metallicity.
In some cases, the spiral structure may extend beyond the core.

The Rayleigh-Taylor (RT) stable boundary between the low entropy phase from below (\ie closer to the center of the cluster) and high entropy from above forms a discontinuity, observed in the form of an X-ray edge known as a cold front (CF) when projection effects are favorable.
Such CFs were found in more than half of the otherwise relaxed CCs \citep{MarkevitchEtAl03Proc}, and nearly in all of the well-observed, low redshift cores.
For example, at least one CF was found in each of the 10 CCs in the sample of \citet{GhizzardiEtAl10}.
These CFs typically show deviations from hydrostatic equilibrium \citep[\eg][]{MazzottaEtAl01}, a metallicity gradient, and a mild pressure jump.
In this work we focus exclusively on such core-type fronts, and not on the CFs found in merger clusters, where the pressure jump is typically large (for a review, see \citet{MarkevitchVikhlinin07}),
nor on the putative CFs created by shock collisions \citep{BirnboimEtAl10}, where no tangential flow, metallicity, or pressure gradients are expected.

Core CFs are often quasi spiral, piecewise spiral, or nearly concentric, suggesting an underlying three dimensional (3D) spiral pattern seen in various projections.
Multiple, quasi-concentric CFs are often found on alternating sides of the center of the cluster, with an increasing distance and size, consistent with a spiral discontinuity manifold observed edge on. Examples include A 2142, RX J1720.1 + 2638, A 2204, and A 496 \cite[][and references therein]{MarkevitchVikhlinin07}.
The thermal profiles across such CFs indicate that they are strongly sheared tangential discontinuities (TDs) seen in projection \citep{KeshetEtAl10}.
These TDs are thought to be stabilized magnetically, with fast, nearly sonic, narrow shear layers lying beneath them \citep{KeshetEtAl10}.

Spiral features and core CFs are typically observed only marginally, at low significance, because a high resolution is needed in order to identify the modest gradients and small TD contrasts in projection, and spectral data are needed to resolve the temperature/metallicity structure. Spirals are rarely unambiguously resolved, but the abundance of subtle spiral features in high quality data suggests the prevalence of spirals, in particular considering projection effects, supporting evidence (such as CFs consistent with edge-on spirals) and the complex core morphologies (hot bubbles, jets, minor mergers, etc.).

Moreover, indirect evidence suggests that CCs have a universal, unrelaxed structure, which we argue is closely related to the spiral features.
First, although CCs differ in their parameters, reside in different gravitational potentials which are strongly affected by the cD galaxies, and are out of equilibrium due to cooling, they broadly adhere to universal thermal and chemical profiles.
Thus, CCs show radial profiles of density, temperature and metallicity that cluster around $\rho\propto r^{-1}$, $T \sim r^{0.4}$, and $Z\sim r^{-0.3}$, respectively \citep[\eg][]{VoigtFabian04, SandersonEtAl06, VikhlininEtAl06, DonahueEtAl06, SandersonEtAl09}.
Second, accumulating evidence indicates that the plasma in many cores, both with and without observed CFs and spiral features, consists of two (or more) spatially separated phases \citep[\eg][]{MolendiPizzolato01, KaastraEtAl04, WernerEtAl10}.
These phases are identified as associated with the two distinct components that alternate to give rise to the spiral structure in select nearby clusters such as Perseus and Virgo, and can be thus interpreted in general, as we discuss in \S\ref{sec:Multiphase} below.

Core CFs and spiral features were modelled as associated with large scale ``sloshing'' oscillations of the ICM, driven by mergers \citep{MarkevitchEtAl01}, possibly involving only a dark matter subhalo \citep{TittleyHenriksen05,AscasibarMarkevitch06}, or by weak shocks/acoustic waves displacing cold central plasma \citep{ChurazovEtAl03, FujitaEtAl04}.
A subset of spiral CFs may be created as low entropy plasma from a subcluster core spirals down the gravitational potential, following an off-axis merger \citep{Clarke04}.

Sloshing simulations have produced CFs and spiral features that nicely resemble observations \citep{AscasibarMarkevitch06, ZuHoneEtAl10, RoedigerEtAl11, ZuHoneEtAl11}.
However, these features are essentially transients \citep{AscasibarMarkevitch06, ZuHoneEtAl10}, which do not survive long once cooling and isolation by the TD-sheared magnetic fields \citep{KeshetEtAl10} are taken into account \citep{ZuHoneEtAl11}.
Moreover, these simulations assume a preexisting CC with the typically observed  size and universal thermal profiles, rather than address the formation of such a core.
Finally, sloshing simulations have not yet addressed the role of AGN feedback and its coupling to the flow.

The preceding discussion suggests that spiral flows are ubiquitous in CCs.
This is based on the combination of the common appearance of spiral structure involving CFs and the interpretation of CFs as tangential bulk flows, is supported by anecdotal evidence such as the intricate velocity structure in Perseus \citep[][figure 8]{SandersEtAl04}, and by the agreement between sloshing simulations and observed spirals.
Moreover, we argue that spiral flows may in fact be essential in the formation of a core, in shaping its (universal) profile, fixing its size, and avoiding catastrophic cooling.
Indeed, this would suggest that CCs are synonymous with spiral flows.

Sloshing simulations have demonstrated that a quasi stable, spiral mode is easily excited in the core, and decays only on a long, $>$Gyr timescale if cooling is not taken into account.
We analytically study spiral modes in the ICM in the presence of cooling and AGN activity.
We constrain the flow only by the observed, nearly spherically symmetric pressure profile, and by the presence of fast flows below spiral CFs inferred in \citet{KeshetEtAl10}.
The implied features of the core are then shown to be strongly constrained, and in good agreement with observations and simulations, even for the simplest two-component toy model we investigate.

The paper is organized as follows.
In \S\ref{sec:Definitions} we introduce the framework and the equations governing the flow.
Power-law scalings of the core and flow parameters are derived in \S\ref{sec:Scalings}.
We study the general properties of the flow in \S\ref{sec:FlowNature}, and explore its two main variants, in which the fast flow beneath CFs is either an inflow or an outflow, in \S\ref{sec:TwoFlowTypes}.
A full solution to one of these variants, which combines a hot inflow and a cold, fast outflow, is derived in \S\ref{sec:Example}.
Heating and feedback are discussed, qualitatively, in \S\ref{sec:Feedback}.
We examine the observational evidence for a multiphase CC plasma in \S\ref{sec:Multiphase}.
Our results are summarized and discussed in \S\ref{sec:Discussion}.

\section{Definitions and governing equations}
\label{sec:Definitions}

The flow is governed by the continuity equation
\begin{equation}
\pr_t \rho + \dvr (\rho \vect{U})=0 \coma
\end{equation}
the momentum equation
\begin{equation}
\left(\pr_t+\vect{U}\cdot \nabla \right)\vect{U}= \vect{a} = -\rho^{-1}\grad{P}+\vect{g} \coma
\end{equation}
and the energy equation
\begin{align}
\rho T\frac{ds}{dt} = \frac{\rho}{\Gamma-1} \left(\pr_t+  \vect{U}\cdot\nabla\right)& \myT + \rho\myT\dvr\vect{U} \\
& =  \dvr(\kappa \grad{T})-\Lambda \nonumber \fin
\end{align}
Here, we denote the mass density $\rho$, the pressure $P$, the temperature $T$, the normalized temperature $\myT\equiv k_B T/\mu$, the Boltzmann constant $k_B$, the mean particle mass $\mu\simeq 0.6m_p$, the proton mass $m_p$, the velocity vector $\vect{U}$, the inertial acceleration vector $\vect{a}$, the entropy $s$, and the thermal conduction coefficient $\kappa$.

For simplicity, the cooling function $\Lambda$ is approximated as
\begin{equation} \label{eq:CoolingFunction}
0<\Lambda\propto \rho^2 T^{1/2} \coma
\end{equation}
independent of metallicity; the effect of a different temperature dependence is discussed at the end of \S\ref{sec:Scalings}.
We approximate the equation of state as that of an ideal gas,
\begin{equation} \label{eq:EOS1}
P = \frac{\rho k_B T}{\mu} = \rho\myT \coma
\end{equation}
with adiabatic index $\Gamma=5/3$.
For simplicity, we assume that the gravitational potential $\Phi$ is spherically symmetric, such that the gravitational acceleration $\vect{g} = - \grad{\Phi}$ is radial.
In addition, we neglect thermal conduction by taking $\kappa=0$.
This is probably justified for conduction perpendicular to the flow, as the shear amplified magnetic fields quench the perpendicular transport.
Parallel conduction is suppressed by scattering off plasma waves \citep[\eg][]{PistinnerEichler98}; the role of parallel conduction is discussed in \S\ref{sec:Feedback}.

The observation of a spiral pattern in a galaxy cluster suggests a symmetry axis, denoted by $z$, perpendicular to the plane of the spiral.
We assume that the spiral flow evolves sufficiently slowly to approach a steady state, in a (possibly rotating) cluster frame of reference.
Therefore, if the spiral pattern is rotating, it must do so uniformly, in order to avoid a winding problem.
Henceforth, we assume that the pattern is rotating with a constant but arbitrary angular frequency $\omega$ about the $z$ axis, and work in the corresponding rotating frame unless otherwise stated.
This requires the addition of fictitious forces to the momentum equation, which then becomes
\begin{eqnarray}
(\vect{U}\cdot \nabla)\vect{U} & = & -\rho^{-1}\nabla \textbf{P} -\nabla\mathbf{\Phi} \\
& & -2\mathbf{\Omega}\times \vect{U} - \mathbf{\Omega}\times(\mathbf{\Omega}\times \vect{x}) \nonumber \coma
\end{eqnarray}
where $\mathbf{\Omega}\equiv \omega \unit{z}$ is the rotation vector.

\begin{figure}[!h]
\centerline{\epsfxsize=7.5cm \epsfbox{\myfig{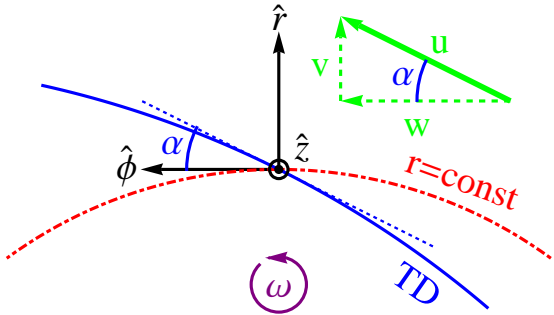}}}
\caption{
Parameter illustration in a plane parallel to the equator.
The TD (solid curve) forms an angle $\alpha$ with $\unit{\phi}$, \ie with the circle $r=\constant$ (dot-dashed).
The plasma must flow parallel to the TD.
The planar velocity $\vect{u}$ (thick green arrow; parallel to the TD at the point analyzed) is decomposed into radial ($\myv$) and angular ($\myw$) components (dashed green arrows).
The pattern may rotate about the $z$-axis, in the $+\unit{z}$ sense if $\omega>0$ (circular arrow).
\label{fig:Defs2}
}
\end{figure}

For simplicity, we assume that an equatorial plane can be found, perpendicular to the $z$ symmetry axis, in which all streamlines are approximately confined to the plane.
We use cylindrical coordinates $\vect{x}\equiv \{r,\phi,z\}$, with $r$ being the radial distance and $\phi$ the azimuth, defined such that the origin coincides with the center of the cluster, and the equatorial plane lies at $z=0$.
We label the velocity components by $\vect{U}\equiv \{\myv,\myw,\myU_z\}$, and sometimes use the two-dimensional (2D) velocity parallel to the equatorial plane, $\vect{u}\equiv \{\myv,\myw\}$.
The TD pattern in such a plane perpendicular to $z$ is quantified by the angle $\alpha(r)$ it makes with respect to $\unit{\phi}$.
These definitions are illustrated in Figure \ref{fig:Defs2}.

The flow along a TD must be parallel to it.
Thus, focusing on the flow along (on either side of) a TD in the equatorial plane, the steady state equations can be cast in a simple form:
continuity,
\begin{eqnarray} \label{eq:Continuity1}
\myprho \equiv \frac{d\ln \rho}{d\ln r} = -\frac{r}{\myv}\vect{\nabla}\cdot \vect{U}
    & = & -\left(1 + \mypw + \mypz \right)
\, ;
\end{eqnarray}
momentum conservation along $\phi$,
\begin{eqnarray} \label{eq:momentum_phi1}
\frac{d}{dr}\left( \frac{\myw^2}{2} \right) + \frac{\myw^2}{r} = \frac{a_\phi}{\gamma} - 2\omega \myw \, ;
\end{eqnarray}
momentum conservation along $r$,
\begin{eqnarray} \label{eq:momentum_r1}
\frac{d}{dr}\left( \frac{\myv^2}{2} \right) - \frac{\myw^2}{r}
= a_r + \omega^2 r+2\omega \myw \, ;
\end{eqnarray}
and energy conservation,
\begin{eqnarray} \label{eq:energy1}
\frac{\myv \rho T}{r} \left(1 + \mypw + \lambda_\tau + \mypz \right) = -\Lambda
\fin
\end{eqnarray}

Here,
$\tau\equiv \myT^{1/(\Gamma-1)}$ is the reduced temperature, $\gamma\equiv \sin{(\alpha)}$ serves as the pattern parameter,
full derivatives are taken along the flow,
\begin{equation} \label{eq:full_der_def1}
\frac{d}{dr} = \frac{\pr}{\pr r} + \frac{1}{\gamma r}\frac{\pr}{\pr\phi} \coma
\end{equation}
and we defined power-law indices
\begin{equation}
\lambda_A\equiv \frac{d\ln |A|}{d\ln r}
\end{equation}
for the evolution of each quantity $A$ along the flow.
These equations apply along streamlines that are parallel to the TD, both above and below it, everywhere that Eq.~(\ref{eq:full_der_def1}) holds.
Deviations are expected far from the TD, but for a spiral flow these are significant only in a localized mixing layer interleaved between the TDs.

\begin{figure}[h!]
\centerline{\epsfxsize=7.5cm \epsfbox{\myfig{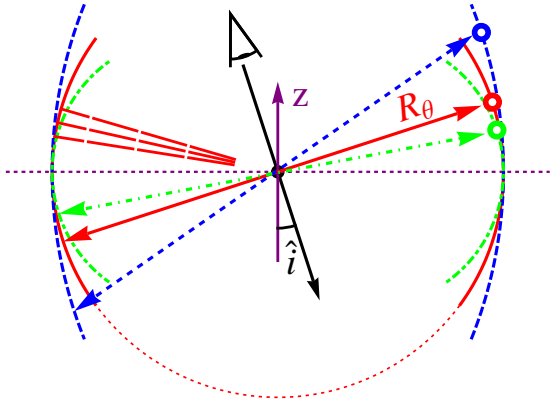}}}
\caption{
Illustration of the flow perpendicular to the equatorial plane, and of CF projection along the line of sight $\unit{i}$ (black arrow).
Three examples of CFs, intersecting the equatorial plane (dotted horizontal line) at the same distance $r$ from the center (black disk), are shown in a plane perpendicular to the equator, \ie ``edge on''.
The CFs have radii of curvature $R_\theta=r$ (solid red), $R_\theta>r$ (dashed blue), and $R_\theta<r$ (dot-dashed green).
The locations (open circles) where the three CFs are parallel to $\unit{i}$ and thus most easily observed in projection, lie at separations (double arrow lines) respectively equal to, longer than, and shorter than $2r$.
The flow is assumed, for simplicity, to be confined to planes (long-dashed red lines shown for the case $R_\theta=r$) spanned by the local TD normal, $\unit{n}$, and $\unit{n}\times\unit{z}$, where $z$ is the symmetry axis (vertical arrow).
\label{fig:Projection}
}
\end{figure}

In order to incorporate the 3D structure of the flow in a simple manner, we assume that $\vect{U}$ and the TD surface are perpendicular when projected onto the $z-r$ plane.
Parameterizing the effects of the flow component perpendicular to the plane of interest is thus simplified, as it depends only on the TD radius of curvature in the $r-z$ plane, $R_\theta=-[1+r'(z)^2]^{3/2}/r''(z)$, through the relation
\begin{equation} \label{eq:mypz_def}
\mypz \equiv r \pr_z \left(\frac{\myU_z}{\myv} \right) = \frac{r}{R_\theta} \fin
\end{equation}
This assumption is illustrated in Figure \ref{fig:Projection}.

Equivalently, this assumption means that the flow is confined to the plane spanned by the local TD normal, $\unit{n}$, and $\unit{n}\times\unit{z}$.
This generalizes the equatorial analysis which follows to any point in the core, both inside and outside the equatorial plane, as long as streamlines within the confined plane remain parallel to the TD.
Note that a flow of this type is well-defined globally only if $r/R_\theta\lesssim 1$, such that the flow planes do not intersect.
Moreover, a pure equatorial spiral pattern is in general deformed in such confined planes that are not parallel to the equator, as $\gamma(r)$ may become complicated and even non-monotonic.

Next, we solve the tangential momentum equation (\ref{eq:momentum_phi1}), by assuming that the inertial acceleration is approximately radial, $|a_\phi|\ll |a_r|$.
This is justified by the approximately spherical distributions of pressure and of total (including dark matter) gravitating mass in CCs, although we shall later see that first order corrections in $|a_\phi/a_r|$ must be retained.
For $a_\phi=0$, Eq.~(\ref{eq:momentum_phi1}) admits two solutions:
\begin{enumerate}
\item
The trivial solution,
\begin{equation} \label{eq:sol_trivial}
\myv = \myw = 0 \coma
\end{equation}
describes corotation with the pattern.
In the inertial (non-rotating) frame, this becomes $\myw^{(i)}=\omega r$.
\item
A non-trivial solution,
\begin{equation} \label{eq:sol_nontrivial}
\myw = \frac{\mycw}{r} - \omega r \, ; \lrgspc \myv = \gamma \left( \frac{\mycw}{r} - \omega r \right) \coma
\end{equation}
where $\mycw$ is a constant.
Here, in the inertial frame, $\myw^{(i)}=\mycw/r$.
For example, for $\mycw=0$, the inertial frame motion is purely radial: the gas propagates inward/outward radially along the rotating spiral, in resemblance of an Archimedes screw.
\end{enumerate}

\section{Scaling of a two-component flow}
\label{sec:Scalings}

An important clue to the nature of core spirals is the presence of fast, nearly sonic flows, found beneath well-observed CFs.
Such flows, and the shear they induce across and below the TDs, are needed \citep[see][]{KeshetEtAl10} in order to explain the thermal profiles observed across CFs, the spatial correlation between CFs and radio minihalos \citep{MazzottaGiacintucci08, KeshetLoeb10}, and the remarkable stability of the TDs, witnessed by the ubiquity, smoothness, and sub-mean free path thinness of the CFs \citep{MarkevitchVikhlinin07}.

\subsection{Nearly adiabatic fast component}

Although the cooling time is shorter than the age of the cluster, it is much longer than the radial sound-crossing time, even deep in the core.
Therefore, the fast flow is approximately adiabatic.
Equations (\ref{eq:EOS1}), (\ref{eq:Continuity1}), and (\ref{eq:energy1}) then yield for this fast (indices $f$) component:
\begin{equation}
\myprho^{(f)} = \frac{\lambda}{\Gamma}=\frac{3}{5}\lambda \simeq -0.36 \coma
\end{equation}
\begin{equation} \label{eq:lambda_Tf}
\mypT^{(f)} = \frac{\Gamma-1}{\Gamma}\lambda=\frac{2}{5}\lambda \simeq -0.24 \coma
\end{equation}
\begin{equation} \label{eq:temp1}
\mypw^{(f)} + \frac{r}{R_\theta} = -\frac{\Gamma+\lambda}{\Gamma}=-1-\frac{3}{5}\lambda \simeq -0.64 \coma
\end{equation}
where we took
\begin{equation}
\lambda\equiv \lambda_P\simeq -0.6
\end{equation}
on the right hand side (RHS) of these equations as a typical value of the logarithmic pressure gradient.
To illustrate the weak dependence upon $\lambda$, Table \ref{tab:ModelSummary} reproduces these and the following relations for a slightly different pressure profile, $\lambda=-2/3$.

Next consider the momentum equation along $\phi$.
The trivial solution Eq.~(\ref{eq:sol_trivial}) corotates with the TD pattern, so the fast component must follow the other, non-trivial solution, Eq.~(\ref{eq:sol_nontrivial}).
Moreover, if the $\omega r$ term in Eq.~(\ref{eq:sol_nontrivial}) were to dominate the fast component anywhere, then Eq.~(\ref{eq:temp1}) would yield $R_\theta<0$, corresponding to outward-curved CFs, unlike the nearly concentric CFs typically observed.
We conclude that the flow beneath the TD must be faster than the pattern throughout the flow, such that Eq.~(\ref{eq:sol_nontrivial}) yields
\begin{equation}
\lambda_{\myw,f} \simeq -1 \coma
\end{equation}
as expected from angular momentum conservation.

Eq.~(\ref{eq:temp1}) now indicates that the perpendicular radius of curvature is larger than $r$,
\begin{equation} \label{eq:CurvatureZ}
\frac{r}{R_\theta} = -\frac{\lambda}{\Gamma} = -\frac{3}{5}\lambda \simeq 0.36 \coma
\end{equation}
so the TDs are intermediate between spherical and cylindrical.
Namely, for an edge-on (\ie within the equatorial plane) observer, the CFs would appear intermediate between concentric semicircles, and lines parallel to the $z$-axis.
Such an intermediate curvature is needed in order to explain the observation of nearly concentric CFs, arguably seen in various projections.
It is also needed for the flow parameterization in Eq.~(\ref{eq:mypz_def}) to be well-defined outside the equatorial plane.
Note that a similar perpendicular TD curvature, of order $r/R_\theta\simeq 1/2$, is found in sloshing simulations \citep{AscasibarMarkevitch06, RoedigerEtAl11}.

\subsection{Radial momentum conservation poorly constrains the power-law model}

The momentum equation in the $r$ direction can be integrated to yield a Bernoulli equation,
\begin{equation} \label{eq:Bernouli1}
\frac{\Gamma}{\Gamma-1}\myT_f + \frac{u_f^2}{2} = C_\Phi(\phi) - \Phi \coma
\end{equation}
where $C_\Phi(\phi)$ is an arbitrary function, $u\equiv |\vect{u}|$, and we neglected the $\omega$-dependent terms by assuming that $|\omega|\ll|\myw|/r$.
However, this equation is not directly useful for deriving the power-law scaling relations.
As the flow is assumed subsonic, the second term on the left hand side (LHS) is subdominant (by at least a factor of three).
As $\Phi$ rises monotonically with $r$ throughout the core's gravitational potential well, Eq.~(\ref{eq:Bernouli1}) is locally consistent with a radially declining temperature profile of the fast flow component, as derived above.
It becomes increasingly difficult to reproduce such a declining temperature profile away from the core, where the gravitational well becomes shallow: this may confine such fast spiral flows to the vicinity of the core.

The slow part of the flow, which strongly dominates the core, is nearly in hydrostatic equilibrium.
For a hydrostatic plasma with a temperature profile $T\propto r^{\mypT\simeq 0.4}$, Eq.~(\ref{eq:Bernouli1}) is thus satisfied for the fast component if\footnote{One may alternatively model the gravitating mass distribution, for example using an NFW profile \citep{NavarroEtAl96} for the dark matter, but the cD galaxy must be included due to its strong influence on the core.}
\begin{equation} \label{eq:C_Phi}
C_\Phi \simeq (-\lambda) T \left(\frac{1}{\mypT}-\frac{1}{\mypT^{(f)}}\right) \simeq 4T \coma
\end{equation}
up to an additive constant.
As $C_{\Phi}$ should be independent of $r$, this calibration holds only locally.
However, the additive constant and the weak radial dependence of $T$ on the RHS of Eq.~(\ref{eq:C_Phi}) indicate that the radial momentum equation can be satisfied globally, by slightly relaxing the approximations leading to Eq.~(\ref{eq:Bernouli1}), including the velocity term in Eq.~(\ref{eq:C_Phi}), or allowing deviations from a pure power-law behavior.

In principle, Eq.~(\ref{eq:Bernouli1}) can be solved for the spiral pattern $\gamma(r)$, as
\begin{equation}
1+\gamma^2 = \frac{2}{\myw_f^2} \left( C_\Phi - \Phi - \frac{\Gamma}{\Gamma-1}\myT_f \right) \fin
\end{equation}
However, the $\gamma$ term is a small correction here ($\gamma^2\simeq 0.04$; see \S\ref{sec:Example}).
Therefore, in addition to the reservations mentioned above, the resulting pattern would be sensitive to the precise form of $\Phi$, would depend on the $\omega$ corrections, and in general would not produce a pure power-law profile.
Instead, we shall compute $\gamma(r)$ using the slow component of the flow.

\subsection{Nearly corotating slow inflow}

Consider the slow flow just above the CF.
If this component were to follow the non-trivial solution Eq.~(\ref{eq:sol_nontrivial}), then there would be fast flows above the CF, at either small or large radii.
Such flows are not observed; rather, the region above the CF is typically consistent with hydrostatic equilibrium \citep{MarkevitchEtAl01, KeshetEtAl10}.
We deduce that the velocity of the slow component is approximately given by the trivial solution Eq.~(\ref{eq:sol_trivial}).
If the flow were adiabatic, this would correspond to stationary, hydrostatic equilibrium plasma in the rotating frame, with an arbitrary thermal profile and consistent with any spiral pattern.

However, a slow flow, in particular one with $\myv\simeq 0$, must be modified by cooling.
Indeed, a stationary flow with $\vect{u}=0$ does not satisfy the energy equation (\ref{eq:energy1}).
Therefore, we must consider small deviations from a trivial flow and from a purely radial acceleration.
Linearizing Eq.~(\ref{eq:momentum_phi1}) to first order in $\{a_\phi,\myv,\myw\}$ yields
\begin{equation} \label{eq:momentum_phi1.5}
2\omega \myv \simeq a_\phi \simeq -\frac{1}{r \rho} \pr_\phi P \coma
\end{equation}
so it is natural to assume the scaling
\begin{equation} \label{eq:momentum_phi2}
\mypv^{(s)} \simeq \mypP - 1 - \myprho^{(s)} \coma
\end{equation}
where indices $s$ pertain to the slow component.
Note that the linearized Eq.~(\ref{eq:momentum_r1}) then implies a small deviation from hydrostatic equilibrium, $a_r^{(s)}\simeq -\omega^2 r+(\rho_s r \gamma)^{-1}\pr_\phi P$.

Finally, combining the continuity equation (\ref{eq:Continuity1}), the tangential momentum equation (\ref{eq:momentum_phi2}), and the energy equation for the cooling, slow component,
\begin{equation}
\mypv^{(s)}+\frac{1}{2} \mypT^{(s)} - 1 \simeq \myprho^{(s)} \coma
\end{equation}
along with the perpendicular radius of curvature determined by the fast component in Eq.~(\ref{eq:CurvatureZ}) and the equation of state (\ref{eq:EOS1}), fixes the scaling of the slow flow:
\begin{equation} \label{eq:SlowScaling1}
\myprho^{(s)} \simeq \frac{-4+3\mypP}{5} \simeq -1.16 \coma
\end{equation}
\begin{equation}
\mypT^{(s)} \simeq \frac{4+2\mypP}{5} \simeq 0.56 \coma
\end{equation}
and
\begin{equation}
\mypv^{(s)} \simeq \frac{-1+2\mypP}{5} \simeq -0.44
\end{equation}
(all three independent of $\Gamma$), and
\begin{equation} \label{eq:SlowScaling4}
\mypw^{(s)} \simeq -\frac{1+3\mypP}{5} + \frac{\mypP}{\Gamma} = -\frac{1}{5}
\end{equation}
(independent of $\mypP$ for $\Gamma=5/3$).

These relations also fix the pattern of the spiral,
\begin{equation}
\mypgamma = \mypv^{(s)}-\mypw^{(s)} \simeq \frac{\Gamma-1}{\Gamma}\mypP = \frac{2}{5}\mypP \simeq -0.24 \fin
\end{equation}
The pattern in turn may be used to determine the scaling of the radial velocity of the fast flow, which was left undetermined above:
\begin{equation}
\mypv^{(f)} \simeq \frac{\Gamma-1}{\Gamma}\mypP -1  = \frac{2}{5}\mypP-1 \simeq -1.24 \fin
\end{equation}

These power-law indices are all fairly weak functions of $\mypP$, indicating that the model is robust.
They are summarized in Table \ref{tab:ModelSummary}, evaluated numerically for a slightly different pressure profile, $\mypP=-2/3$.

\begin{table}[h]
\caption{\label{tab:ModelSummary} Scaling of the two-component model, for $\Gamma=\frac{5}{3}$; $\mypP=-\frac{2}{3}$.
}
\begin{tabular}{l|ll}
\hline
Property $A$ & \hspace{0.3cm} Power law index $\lambda_A=d\ln|A|/d\ln(r)$ \hspace{-4.3cm} & \\
\hline
pressure $P$ & \hspace{1.6cm} $\mypP\equiv \lambda_P =-\frac{2}{3}$ \hspace{-4.3cm}   & \\
\vspace{0.5mm}
spiral slope $\gamma$ & \hspace{1.3cm} $\frac{2}{5}\mypP = -\frac{4}{15} \simeq -0.27$ \hspace{-1.3cm}   &   \\
\hline
\vspace{0.5mm}
 & Fast phase & Slow phase \\
\hline
density $\rho$ & $\frac{3}{5}\mypP=-
\frac{2}{5}=-0.4$    & $\frac{-4+3\mypP}{5}= -\frac{6}{5} = -1.2$ \\
\vspace{0.5mm}
temperature $T$ & $\frac{2}{5}\lambda = -\frac{4}{15}\simeq -0.27$     & $\frac{4+2\mypP}{5} = \frac{8}{15} \simeq 0.53$ \\
\vspace{0.5mm}
rad. velocity $\myv$ & $\frac{2}{5}\mypP-1 = -\frac{19}{15} \simeq -1.27 $    & $\frac{-1+2\mypP}{5} = -\frac{7}{15} \simeq -0.47$ \\
tan. velocity $\myw$ & $-1$    & $-\frac{1}{5}=-0.2$ \\
\hline
\end{tabular}
\footnotetext{
Note that $\mypP$ here is taken slightly different numerically than in the text, in order to show its rather weak effect, in particular on the slow component.
The flow equations require $r/R_\theta = -\mypP/\Gamma = 2/5$, intermediate between cylindrical and spherical TD geometry.
}
\end{table}

The above scaling relations imply that the bracketed term in the energy equation (\ref{eq:energy1}) is positive for the slow component,
\begin{equation} \nonumber
1 + \mypw^{(s)} + \lambda_\tau^{(s)} + \mypz \simeq [4\Gamma+(5-3\Gamma)\mypP]/[5(\Gamma-1)]\simeq 2 \fin
\end{equation}
Therefore, the slow component must be an inflow, with $\myv_s<0$.
Approximating the cooling rate as $\Lambda \simeq 2.1 \times 10^{-27}n[\mbox{cm}^{-3}]^2 T[K]^{1/2}\erg\se ^{-1}\cm^{-3}$ \citep{RybickiLightman86}, we find
\begin{equation} \label{eq:cooling_velocity}
\myv_s \simeq -10 \, n_{0.03} T_4^{-1/2} r_{10} \km \se^{-1} \coma
\end{equation}
where $n_{0.03}$ is the electron number density $n=\rho/\mu$ in $0.03\cm^{-3}$ units, $T_4 \equiv (k_B T/4\keV)$, and $r_{10}\equiv (r/10\kpc)$. This velocity is of order $1\%$ of the sound velocity $c_s$, consistent with the slow flows above CFs inferred from observations.

Notice that the spiral pattern and fast flow are independent of, and the slow flow depends weakly on, the precise temperature dependence of the cooling function Eq.~(\ref{eq:CoolingFunction}).
For example, in the cold, high metallicity regions in the centers of small clusters and groups of galaxies, the temperature dependence weakens to $\Lambda \propto T^0$ \citep[\eg][]{PetersonFabian06}.
Here, the scaling laws in Eqs.~(\ref{eq:SlowScaling1})--(\ref{eq:SlowScaling4}) slightly change, to
$\myprho^{(s)} \simeq -1.07$,
$\mypT^{(s)} \simeq 0.47$,
$\mypv^{(s)} \simeq -0.53$,
and
$\mypw^{(s)} \simeq -0.29$.

\section{General properties of the flow}
\label{sec:FlowNature}

The scaling relations derived in \S\ref{sec:Scalings} indicate that the velocity of the fast, cold component increases at smaller radii, while the slow, hot flow becomes denser and somewhat cooler at small radii.
The spiral pattern is found to be nearly logarithmic (for a logarithmic spiral $\gamma=\const$, or equivalently, $\mypgamma= 0$), with a slight tendency towards an Archimedes spiral (for which $\mypgamma=-1$).
We may now combine these scaling laws with observations in order to test the model and further explore the nature of the flow.

Observations suggest \citep{KeshetEtAl10} that the fast flow inward of a CF is confined to the vicinity of the discontinuity, inducing shear up to distances of order $\Delta r \lesssim 0.2 r$ below the CF at radius $r$.
Therefore, in the simple approximation where the core is composed of only the two, fast and slow, components, we expect the thermal properties of the core to be dominated by the slow flow.
Hence, the core must approximately follow $\rho\sim r^{-1}$ and $T\sim r^{1/2}$.
Similarly, the behavior of the TD shear is dominated by the fast flow, so approximately $\Delta u\equiv u_s-u_f \sim r^{-1}$.

\begin{figure*}[!ht]
\centerline{\epsfxsize=9.5cm \epsfbox{\myfig{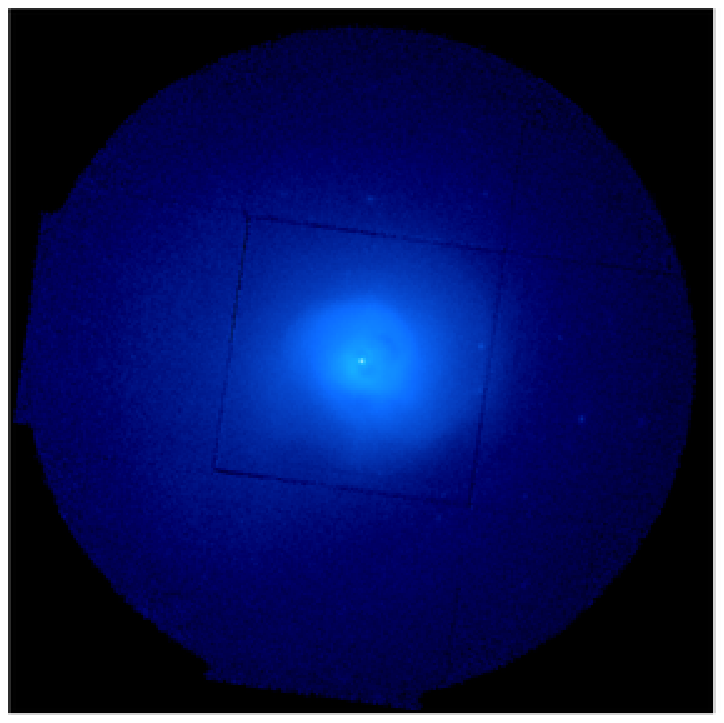}}
\epsfxsize=7cm \epsfbox{\myfig{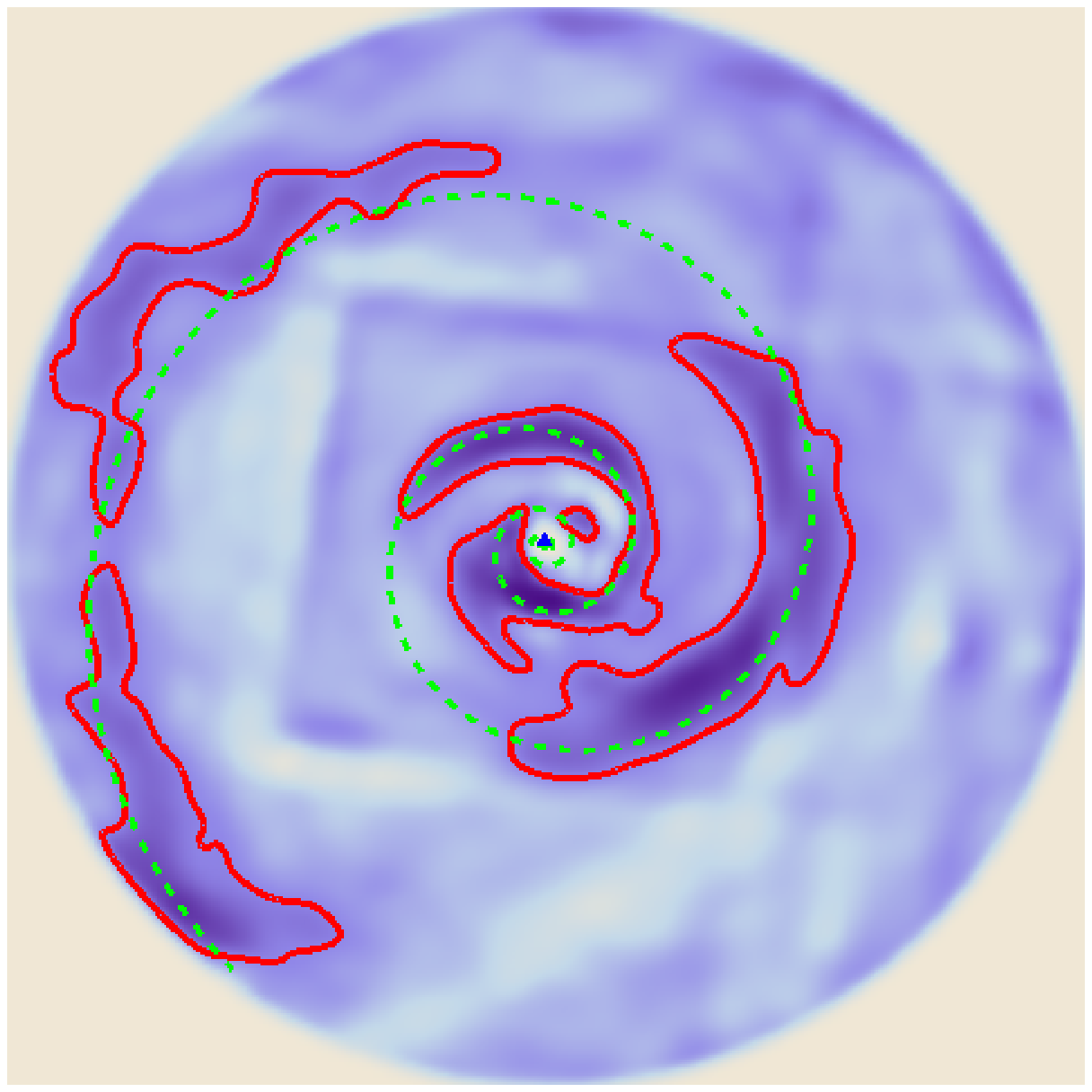}}}
\caption{
Spiral pattern observed in Perseus.
Shown is an XMM image \citep[][data courtesy: E. Churazov]{ChurazovEtAl03}, before (left; $30'$ diameter) and after (right; $25'$ diameter) smoothing with an adaptive, photon count per-bin conserving, Gaussian filter, cropping edge effects, and applying a gradient filter.
Red contours enclose dark regions that correspond to enhanced gradients, tracing CFs.
The light regions have small gradients, and show mosaic and other artifacts.
The dashed green curve is the best-fit power-law spiral derived in Figure \ref{fig:PerseusPattern2}.
\label{fig:PerseusPattern1}
}
\end{figure*}

\begin{figure}[h]
\centerline{\epsfxsize=8cm \epsfbox{\myfig{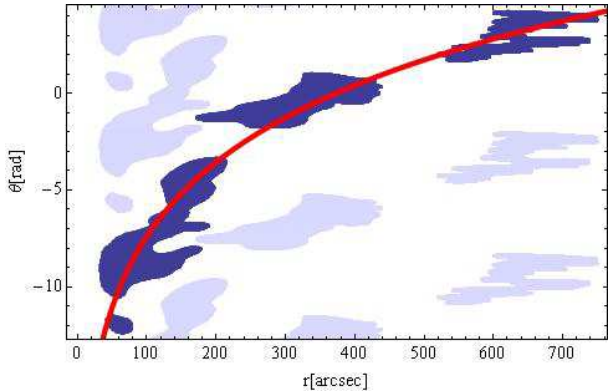}}}
\caption{
Discontinuity pattern in Perseus.
The strong gradient regions (enclosed by red contours in Figure \ref{fig:PerseusPattern1}) are shown in the $r$--$\theta$ phase space.
Although $\theta$ is $2\pi$-periodic (pale regions), one may trace it along an extended interval (dark regions) by following the pattern in Figure \ref{fig:PerseusPattern1} and varying the threshold.
The best fit (red curve) power-law spiral, $\theta[\mbox{rad}]\simeq -107.8 + 78.3 (r/\mbox{arcsec})^{0.054}$, corresponds to $\lambda_\gamma=-1-r \theta''(r)/\theta'(r) \simeq -0.05$: nearly a logarithmic spiral.
\label{fig:PerseusPattern2}
}
\end{figure}

\begin{figure}[!h]
\centerline{\epsfxsize=7cm \epsfbox{\myfig{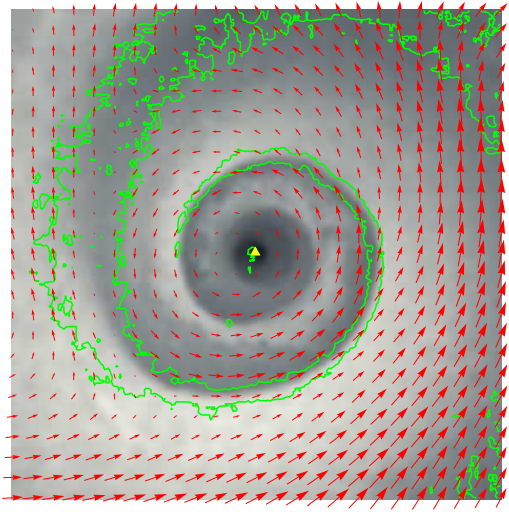}}}
\caption{
Spiral flow formed as a result of a mass ratio 5 merger with a dark matter subcluster, in the adiabatic SPH simulation of \citet[][data courtesy: Y. Ascasibar]{AscasibarMarkevitch06}.
The slice shown is $560\kpc$ on the side.
Greyscale shows the gas temperature, ranging from $3.3\keV$ (dark) to $9.8\keV$ (bright).
Arrows trace the gas flow.
Green contours enclose regions in which $(\crl\vect{U})_z$ is smaller than a negative threshold, \ie strong local clockwise circulation.
The dark matter peak is shown as a yellow triangle.
The spiral pattern rotates counterclockwise, \ie is trailing.
The fast component here is an inflow.
The slow component is in part an outflow, although this is obscured by the uniform rotation of the pattern.
\label{fig:SPHPattern1}
}
\end{figure}

\begin{figure}[t!]
\centerline{\epsfxsize=8cm \epsfbox{\myfig{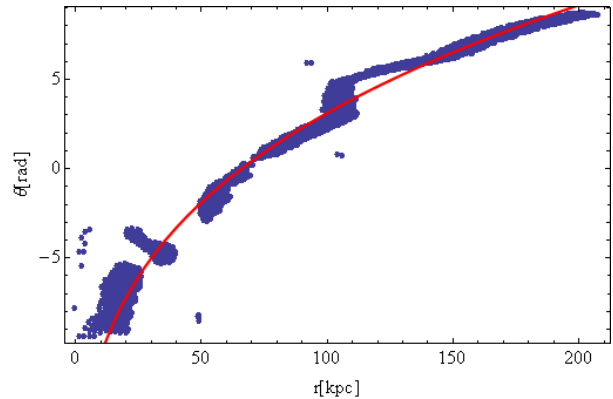}}}
\caption{
Discontinuity pattern in the simulation of \citet{AscasibarMarkevitch06} shown in Figure \ref{fig:SPHPattern1}, derived in the same method as Figure \ref{fig:PerseusPattern2}, based on regions exceeding a gradient threshold (blue points).
The best power-law spiral fit (red curve), $\theta=-26.2 + 8.5 (r/\mbox{kpc})^{0.27}$, corresponds to $\lambda_\gamma\simeq -0.27$: nearly a logarithmic (slightly tending towards an Archimedes) spiral.
\label{fig:SPHPattern2}
}
\end{figure}

These properties are broadly consistent with the thermal structure of observed cores \citep[\eg][]{VoigtFabian04, SandersonEtAl06, VikhlininEtAl06, DonahueEtAl06, SandersonEtAl09}, and with the (presently weak) constraints on shear \citep{KeshetEtAl10}.
We also find that the spiral pattern inferred above is consistent with the observed spiral pattern in Perseus, which is amongst the clearest spirals observed to date, and is presumably seen nearly face-on (approximately along the $z$-axis); see Figures \ref{fig:PerseusPattern1} and \ref{fig:PerseusPattern2}.
Even better agreement is found with the late-time, quasi-stable pattern discovered in the merger simulation of \citet{AscasibarMarkevitch06}, where the projection is controlled; see Figures \ref{fig:SPHPattern1} and \ref{fig:SPHPattern2}.
In both cases, the spiral is approximately logarithmic, with slightly negative $\mypgamma$.

The two components of the flow slide across each other along the spiral, RT stable, TD surface that separates between the fast, cold flow from below and the slow, hot component from above.
This strongly sheared boundary is thought to be stabilized magnetically, isolating the two components \citep{KeshetEtAl10}.
In contrast, the transition layer between slow flow from below and fast flow from above is not discontinuous, as such a discontinuity would become RT unstable.
This layer is likely to involve substantial heat conduction, convection, and gas mixing, and may become turbulent.
In this region our analysis breaks down, as streamlines may no longer be parallel to the TD.

In order for the TD to remain RT stable, the fast flow below the discontinuity must remain at least as dense and as cool as the flow above the TD.
The temperature (or equivalently, as we assume an isobaric transition, the density) contrast $q$ across the TD increases with $r$ as
\begin{equation} \label{eq:TRatio}
1 < q \equiv \frac{T_s}{T_f} \simeq \frac{\rho_f}{\rho_s}\propto r^{4/5} \coma
\end{equation}
so the discontinuity would become RT unstable at very small radii, limiting the extent of the flow or the validity of the scaling derived above.
Similarly, the velocity ratio between the fast and slow components diminishes with increasing radius,
\begin{equation} \label{eq:vRatio}
1< \frac{u_f}{u_s} \simeq \mp \frac{\vect{u}_f}{\vect{u}_s} \propto r^{-4/5} \coma
\end{equation}
where the minus (plus) sign corresponds to a fast outflow (inflow).
Thus, the shear responsible for the stabilizing magnetic fields gradually diminishes at large radii, again limiting the extent of the flow or the validity of the power-law scaling.
Note that the $\pm4/5$ power-law indices in Eqs.~(\ref{eq:TRatio}) and (\ref{eq:vRatio}) are independent of $\Gamma$ and $\mypP$.

As TD contrasts of order $q\sim 2$ and nearly sonic shear $\Delta u\sim c_s$ are inferred from CF observations, extrapolating these scalings to small radii suggests that the two components have similar temperatures and densities at some small radius $r_0$, near the base of the TD, where $q$ approaches unity and the fast component is nearly sonic,
\begin{equation} \label{eq:SpiralBase}
q(r_0)\simeq 1 \smlspc ; \lrgspc  u_f(r_0)\simeq c_s \fin
\end{equation}

More generally, a measured contrast $q(r)$ or shear $\Delta u(r)$ thus places an upper limit on the radial extent of the discontinuity, for example $r_0>q(r)^{-5/4}r$.
As an illustration, a TD following the power-law scalings of \S\ref{sec:Scalings} extends no more than a factor $r/r_0\simeq 2.4$ ($\simeq 3.9$) below a CF observed at radius $r$, if the inferred TD contrast there is $q(r)=2$ ($=3$).
In practice, the flow may extend somewhat beyond such estimates, because \myNi the flow can persist beyond the TD, if the discontinuity evolves into a more gradual transition; \myNii inaccurate deprojection may cause $q$ and $\Delta u$ to appear smaller than they really are, for example if a very dense or fast flow is confined to the very near vicinity of the TD; and \myNiii deviations from our power-law model are expected at small radii, due to the nearly sonic velocity of the fast component, the possible breakdown of the assumption $u_s\equiv |\vect{u}_s|\ll\omega r$, and the active environment near the AGN.

The scaling relations derived in \S\ref{sec:Scalings} are likely to break down at some large radius $r_{\mymax}$, where the fast flow becomes sufficiently slow that its cooling can no longer be neglected.
We may use Eqs.~(\ref{eq:cooling_velocity}), (\ref{eq:vRatio}), and (\ref{eq:SpiralBase}), to crudely estimate that
\begin{equation} \label{eq:r_max}
r_{\mymax} \simeq \left(\frac{c_s}{|\myv_s|} \right)_{r_0}^{1/2} \simeq 100 \left(\frac{r_{10} T_4}{n_{0.03}}\right)_{r_0}^{1/2} \kpc \fin
\end{equation}
The spiral flow may persist beyond $r_{\mymax}$ with a somewhat different scaling, or may retain the same scaling if the cooling of the cold component is compensated, for example by heat conduction from the hot component.
Note that by $r_{\mymax}$, the contrast across the TD becomes large, of order $q\simeq (r_{\mymax}/r_0)^{4/5}\simeq 6$.

Alternatively, the TD may be disrupted by Kelvin-Helmholtz (KH) and other instabilities near or somewhat above $r_{\mymax}$.
At small radii, shear magnetic amplification is thought to generate strong magnetic fields that stabilize the discontinuity \citep{KeshetEtAl10}. At larger radii, the shear gradually weakens. This implies weaker KH instabilities, on one hand, but slower magnetic amplification, on the other. Taking into account additional (\eg Richtmyer--Meshkov) instabilities, turbulence, and the weaker seed magnetic fields anticipated at large radii, this suggests that the two components mix at some radius $r\gtrsim r_{\mymax}$.
Both effects may contribute to the absence of observed CFs with large contrasts.

The momentum equations for the slow component require that $\omega\neq 0$, \ie the spiral pattern must rotate, albeit much slower than the fast component.
The sense of rotation is related to the tangential force acting on the slow component, by Eq.~(\ref{eq:momentum_phi1.5}) (with $\myv<0$): if $a_\phi$ points towards the outspiral (inspiral) direction, then the spiral is trailing (leading).
For a spherically symmetric gravitational potential, $\sign(a_\phi)=-\sign(\pr_\phi P)$, so the sense of rotation is fixed by the minute tangential pressure gradients observed.

In a quasi spherical system, the deviation of the pressure map from its azimuthal average may reveal such small gradients.
In Perseus, such a procedure shows a slightly elevated thermal pressure significantly outside the CFs, at a finite distance above them: compare the pressure map in figure 10 of \citet{ChurazovEtAl03} to the temperature or entropy maps.
This corresponds to a pressure gradient within the slow component pointing towards the inspiral direction.
It suggests that the spiral pattern is \myNi indeed rotating; and \myNii trailing, because the force $\sim a_\phi$ exerted on the slow component is directed along the outspiral direction.
The above conclusions are not sensitive to the presence of a magnetic pressure, because the elevated thermal pressure is found significantly far above the CF: one does not expect the magnetic pressure to increase to significant levels as one approaches the CF from above.

The projected appearance of a spiral TD depends on the orientation of the spiral axis with respect to the line of sight, and on the 3D structure of the TD manifold.
As CFs are often nearly concentric, and some spiral CFs are observed, $R_\theta/r$ cannot be much larger or much smaller than unity.
Some projection effects are illustrated in Figure \ref{fig:Projection} (a full analysis of projection is deferred to future work).
The figure shows how, for $R_\theta>r$ (in dashed blue), as the line of sight gradually deviates from face-on viewing (in which the line of sight is parallel to the $z$ axis of symmetry), the spiral pattern is stretched \emph{parallel} to the projected $z$ axis.
An opposite, perpendicular stretching occurs for $R_\theta<r$ (shown in dot-dashed green).
Due to the poor statistics, it is difficult to measure this curvature observationally and test Eq.~(\ref{eq:CurvatureZ}) and the corresponding parallel stretching.

We have determined the radius of curvature of the flow perpendicular to the equatorial plane, but did not address the evolution of streamlines far from the equator, nor have we specified the extent of the spiral flow above the plane. The flow must extend a substantial fraction of the way towards the poles, in order to account for the observation of spiral patterns at various projections.
In principle, the flow can extend all the way to the pole, where the model streamlines have an approximate helical pattern around the $z$-axis, with an amplitude that diminishes close to the pole.

\section{Two types of composite flows}
\label{sec:TwoFlowTypes}

Due to the adiabatic nature of the fast component, its behavior is symmetric under time-reversal.
Hence, in contrast to the slow component which must be an inflow\footnote{This is due to cooling. A slow outflow is possible, for
example, in an adiabatic simulation.}, the analysis above does not distinguish whether the fast component is an inflow or an outflow.
We thus consider below both options for the composite spiral flow: \myNi a combination of slow and fast inflows; or \myNii a combination of a fast outflow and a slow inflow.

The former type of flow is essentially a transient that accretes mass onto the center, while the latter could in principle be long-lived, and may form a closed circulation loop with quenched accretion.
Note that although spiral inflow and outflow components carry opposite signs of mass and heat radial fluxes, they carry the same sign of angular momentum flux.
Hence, both an inflow and an outflow similarly transfer angular momentum in the trailing (leading) spiral sense inward (outward), across the core, limiting the lifetime of the flow unless an external torque is involved.

\subsection{fast and slow inspiral}
\label{sec:InflowInflow}

The first option, in which the fast component is an inflow, is in essence a transient phenomenon (albeit long-lasting, of timescale $\gtrsim \mbox{Gyr}$ in the absence of cooling), as the cold, metal-rich inflow must originate deep in the core.
In order to produce such a flow, the central plasma must first be uplifted by some violent process, such as the gravitational kick and angular momentum deposition imparted by a merger event \citep{MarkevitchEtAl01}, the impact of weak shocks or acoustic waves \citep{ChurazovEtAl03, FujitaEtAl04}, or the uplift of hot (radio) AGN bubbles that buoyantly rise from the center.

In particular, sloshing motions induced by strong perturbations of the core following a merger event were shown to produce CFs and spiral patterns in broad agreement with observations, provided that the merger does not strongly disrupt the original core.
This was demonstrated in simulations of major mergers with a dark matter subhalo \citep{TittleyHenriksen05, AscasibarMarkevitch06} and of minor mergers \citep{RoedigerEtAl11}, and in simulations that incorporated cooling \citep{ZuHoneEtAl10, ZuHoneEtAl11}.

However, without additional effects, such sloshing cannot protect CCs from catastrophic cooling.
Although shown to somewhat delay cooling by mixing the gas \citep{ZuHoneEtAl10}, sloshing alone cannot sustain the core for more than $\sim 1 \Gyr$, once magnetic shear amplification along the TDs \citep{KeshetEtAl10} is taken into account \citep{ZuHoneEtAl11}.
Nevertheless, it remains to be seen if AGN feedback could stem cooling in the presence of such a composite spiral inflow.

\begin{figure*}
\centerline{\epsfxsize=7cm \epsfbox{\myfig{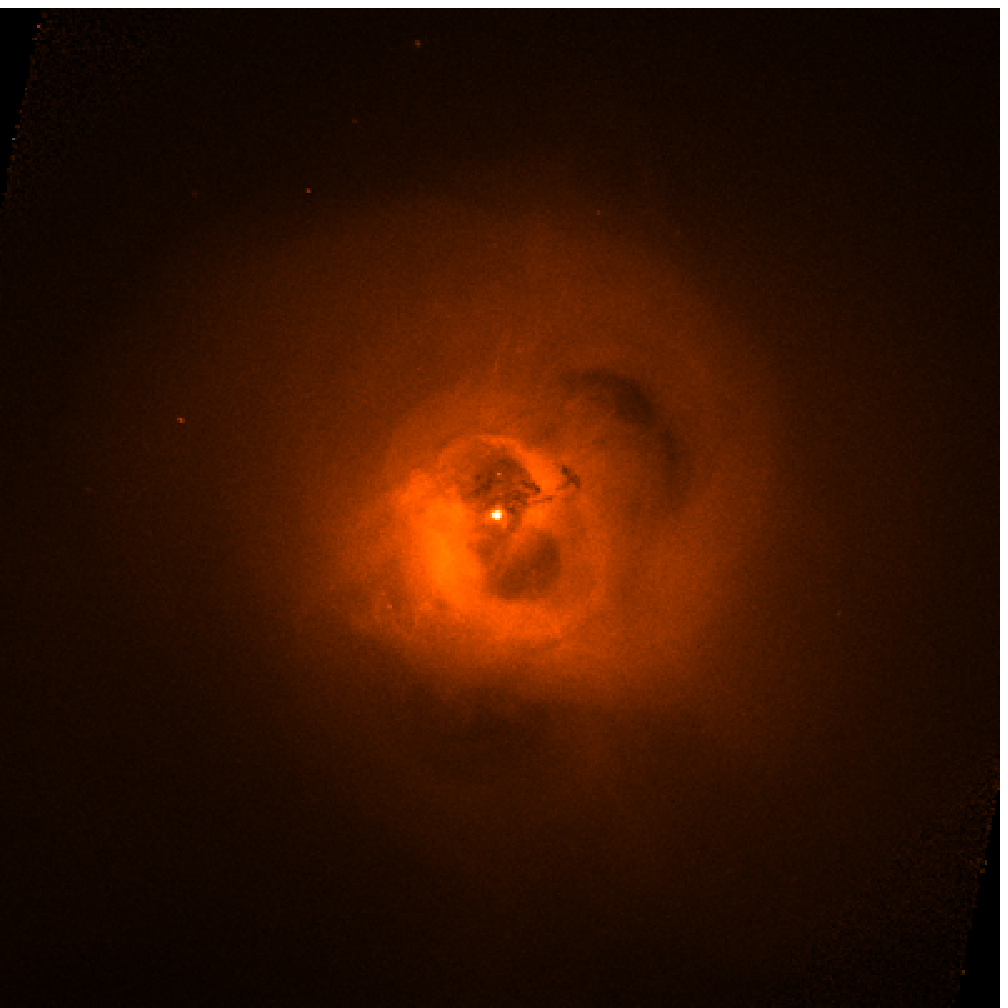}}
\epsfxsize=7cm \epsfbox{\myfig{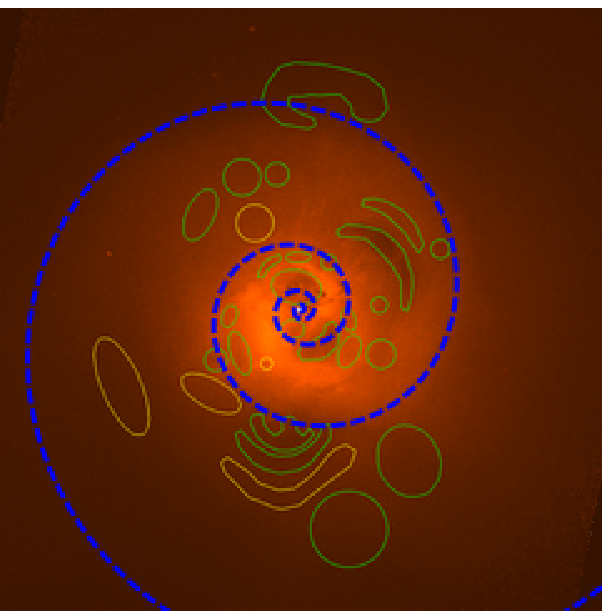}}
}
\caption{
Deep ($900$ ks ACIS-S) archival Chandra image of Perseus \citep[presented, \eg by][]{FabianEtAl06, SandersFabian07}, $8'$ ($\sim240\kpc$) on the side, shown with (right) and without (left) the features discussed in the text.
Over-plotted on the right are the best-fit spiral CF pattern (dashed blue) derived in Figure \ref{fig:PerseusPattern2}, along with apparent (green closed curves) and possible (yellow curves) X-ray cavities, suggesting AGN bubbles interacting, and possibly entrained in the spiral flow.
The X-ray cavities were identified by a visual inspection (M. Markevitch, private communications, 2009, 2011);
some of them were also pointed out in \citet{FabianEtAl11}.
\label{fig:PerseusBubbles}
}
\end{figure*}

Strong magnetic fields, which are thought to form not only at the TD, but throughout the fast flow layer \citep{KeshetEtAl10}, as recently demonstrated numerically \citep{ZuHoneEtAl11}, can divert AGN bubbles and entrain them within the spiral pattern.
Such entrainment is indeed observed in some cores, as illustrated for Perseus in Figure \ref{fig:PerseusBubbles}. See also \citet{FormanEtAl07, FabianEtAl11}.
Projection effects confuse the interpretation of such images, in particular in regards to the interaction of the spiral flow with entrained bubbles, linear cold filaments, and other localized features.
For example, some bubbles axes or long filaments may in fact be oriented perpendicular to the equatorial plane, along the $z$ symmetry axis.

In addition to enhancing the dissipation of bubble energy, such interaction between rising bubbles and a spiral inflow suggests a natural feedback mechanism: accretion-driven AGN activity may disrupt or heat the inflow.
Another possibility is that mechanical feedback may stop the inflow and push it outwards, resulting in an inflow--outflow composite spiral of the type studied in \S\ref{sec:OutflowInflow} below.
Feedback is further discussed in \S\ref{sec:Feedback}.

\subsection{fast outspiral and slow inspiral}
\label{sec:OutflowInflow}

The second possible type of composite flow combines a fast, cold \emph{outflow} and a slow, hot inflow, spiraling across each other.
Such a flow follows the same scaling relations as the double-inflow model of \S\ref{sec:InflowInflow}, although shear and gas mixing are probably somewhat stronger.

It is interesting to note that adiabatic sloshing simulations sometimes show, at least locally, composite flows of an opposite character, combining a fast inflow and a slow outflow, as seen for example in some regions in the snapshot of the \citet{AscasibarMarkevitch06} simulation depicted in Figure \ref{fig:SPHPattern1}.
The flow discussed here is similar, in the adiabatic limit, to the time-reversal of such an adiabatic simulation.

A novel property of a composite inspiral--outspiral flow of the type discussed here is that, unlike the composite inflow model of \S\ref{sec:InflowInflow}, it could constitute a quasi steady-state solution, as long as the center can absorb angular momentum in the trailing spiral sense.
This can occur, for example, if an external force torques the center in the opposite, leading spiral direction.
Feedback-regulated flows that alternate between fast and slow mass accretion, or between spirals of an opposite orientation, are also possible.

Specifically, consider the mass flux across a given radius in the equatorial plane.
The outflow layer is much faster than the inflow, but is also much narrower, so the relative widths of the two components (defined, say, as the ratio $\nu(r)\equiv \Delta \phi_f/\Delta \phi_s<1$ of their $\phi$ extents at a given $r$) can be chosen to eliminate the radial mass flux.
Due to the curvature of the streamlines both inside and outside the equatorial plane, the planar mass flux $\rho \myv r$ of each component is $r$-dependent. Nevertheless, for both components it follows the same scaling,
\begin{equation} \label{eq:jScaling}
\frac{d\ln(\rho \myv r)}{d\ln r} = \mypgamma + \frac{r}{R_\theta} = \mypP \fin
\end{equation}
Thus, a fixed ratio $\nu$ between inflow and outflow volume fractions would guarantee a choked mass accretion throughout the flow, if $\nu$ is properly tuned.
Note that, as the streamlines are approximately parallel, such a spiral flow carries no angular momentum in the rotating reference frame, although it does transfer angular momentum radially, across the flow region.

Such a non-accreting flow essentially constitutes a closed circulation loop, which could explain the low star formation rate observed in cores \citep[\eg][]{RaffertyEtAl08, ODeaEtAl08}.
Here, cold gas from the base of the flow at $r_0$ is rapidly removed from the center outwards, decelerates, and in part slowly accretes back to the center, after being non-adiabatically heated by close proximity to hotter gas, through gas mixing and heat conduction.
The flow thus solves the cooling problem, by advecting heat from the huge heat reservoir outside the core and removing dense gas from the center before it can catastrophically cool, while quenching or maintaining a low level of mass accretion, and carrying angular momentum away from the center.

Such a flow requires some mechanism to rapidly propel gas from the central parts of the core outwards.
Candidate mechanisms include gravitational forces, such as the stirring effect of the rotating vector separating between gas and dark matter peaks, and the AGN energy output, coupled for example through the forces exerted by shocks, jets, and buoyant hot bubbles.
Such AGN bubbles, in particular, suggest a natural feedback mechanism, provided that their production is moderated by the accretion rate, as discussed in \S\ref{sec:Feedback}.

High angular momentum in the inner core is another requirement, needed both in order to sustain the angular momentum flux, and to loosen the gravitational binding of the core.
Such angular momentum can build up through (possibly alternating) epochs of strong spiral flows, for example in the form of an inspiral--inspiral composite flow, by merger events, and by the sinking of dark matter substructure towards the center.
The initial formation of the spiral pattern can be facilitated by a merger event, by amplified g-modes, or by spiral instabilities that tend to separate even a non-rotating, uniformly accreting, cooling gas into spiral motions of opposite orientation, for example in the standing accretion shock instability (SASI) simulations discussed in \citet{BlondinShaw07}.

\section{Two-component inflow--outflow toy model}
\label{sec:Example}

Above we derived some of the properties and scaling relations of a spiral flow of the type expected in a cluster core, constrained by the presence of a spiral TD surface with strong shear due to a fast flow below the discontinuity.
We showed that the flows immediately above and immediately below the TD follow different scaling relations, but have not quantified the variation in the flow properties with distance from the TD.
Here we wish to provide the full solution to one type of such a spiral flow, by introducing a simple two-component toy model.

In order to obtain a fully constrained system and reproduce the low mass accretion rate inferred from observations, we choose the flow variant of \S\ref{sec:OutflowInflow}, combining a fast outflow and a slow inflow, and require that the radial mass flux vanishes.
In order to account for the mild, $q\sim 2$ contrast of typical CFs, and the nearly sonic shear inferred from their profiles, we assume that $\rho_f=\rho_s$, $T_f=T_s$, and $u_f\equiv |\vect{u}_f|=c_s$ at some small radius $r_0$, as discussed in \S\ref{sec:FlowNature}; see Eq.~(\ref{eq:SpiralBase}).
These constraints fix all the model parameters except the normalization of $\gamma$, which can be approximately measured as demonstrated below.

The requirement of choked mass accretion fixes the ratio between the widths of the two layers, which we parameterize using the angular extents $\Delta \phi$ of each phase at a given $r$,
\begin{equation} \label{eq:nu}
\nu \equiv \frac{\Delta \phi_f}{\Delta \phi_s} \simeq \frac{\rho_s \myv_s}{\rho_f \myv_f} \Big|_{r_0} \simeq
\frac{\myv_s}{c_s \gamma} \Big|_{r_0} \simeq 0.05 \left(\frac{n_{0.03} r_{10}}{T_4 \gamma_{0.2}}\right)_{r_0} \fin
\end{equation}
Here we defined $\gamma_{0.2}\equiv \gamma(r_0)/0.2\simeq 1$, based on preliminary estimates of $\gamma$ at the base of the spiral patterns:
\myNi in Perseus, where $\gamma(10\kpc)\simeq 0.20$ (see Figure \ref{fig:PerseusPattern1}); and \myNii in the \citet{AscasibarMarkevitch06} simulation, where $\gamma(10\kpc)\simeq 0.23$ (see Figure \ref{fig:SPHPattern1}).
Recall that although $\nu$ is evaluated at $r_0$, it remains constant along the spiral; \cf Eq.~(\ref{eq:jScaling}).

A logarithmic spiral follows $r(\phi)\propto e^{\gamma \phi}$.
As the spiral patterns observed are nearly logarithmic, the radial separation between consecutive TD spiral arms in the equatorial plane is approximately $r_{n+1}/r_n \simeq e^{2\pi \gamma} \simeq  3.5 e^{2\pi(\gamma-0.2)}$.
In the toy model, the fast flow is confined to a thin layer below the TD, with an approximate radial thickness of
\begin{equation}
\Delta r_f \simeq r\left( 1-e^{-2\pi\gamma \nu}\right) \simeq 0.06r \coma
\end{equation}
where we used the parameters of Eq.~(\ref{eq:nu}).
In reality, the fast layer may be more extended, and the transition between the fast and slow phases must be gradual.

Next, we compute the azimuthal averages of the flow properties within the equatorial plane.
Note that although these quantities approximately give the projected radial profiles when the streamlines lie on spheres (\ie when $R_\theta\sim r$), a viewing angle-dependent correction is needed to account for projection effects when $R_\theta\neq r$, in particular when the spiral is not observed face-on.
Nevertheless, as long as $R_\theta/r$ is not too far removed from unity, such corrections are in general small.

\begin{figure*}[ht]
\centerline{\epsfxsize=8cm \epsfbox{\myfig{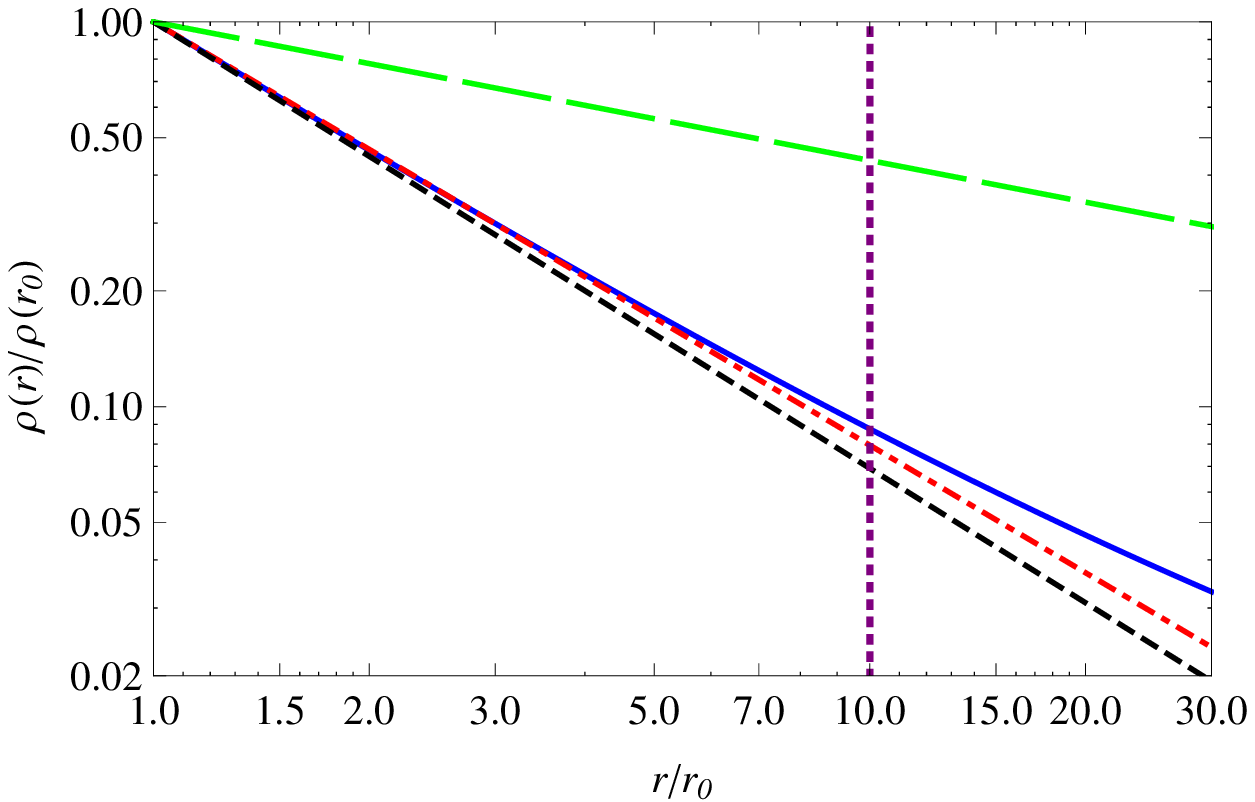}}
\epsfxsize=8cm \epsfbox{\myfig{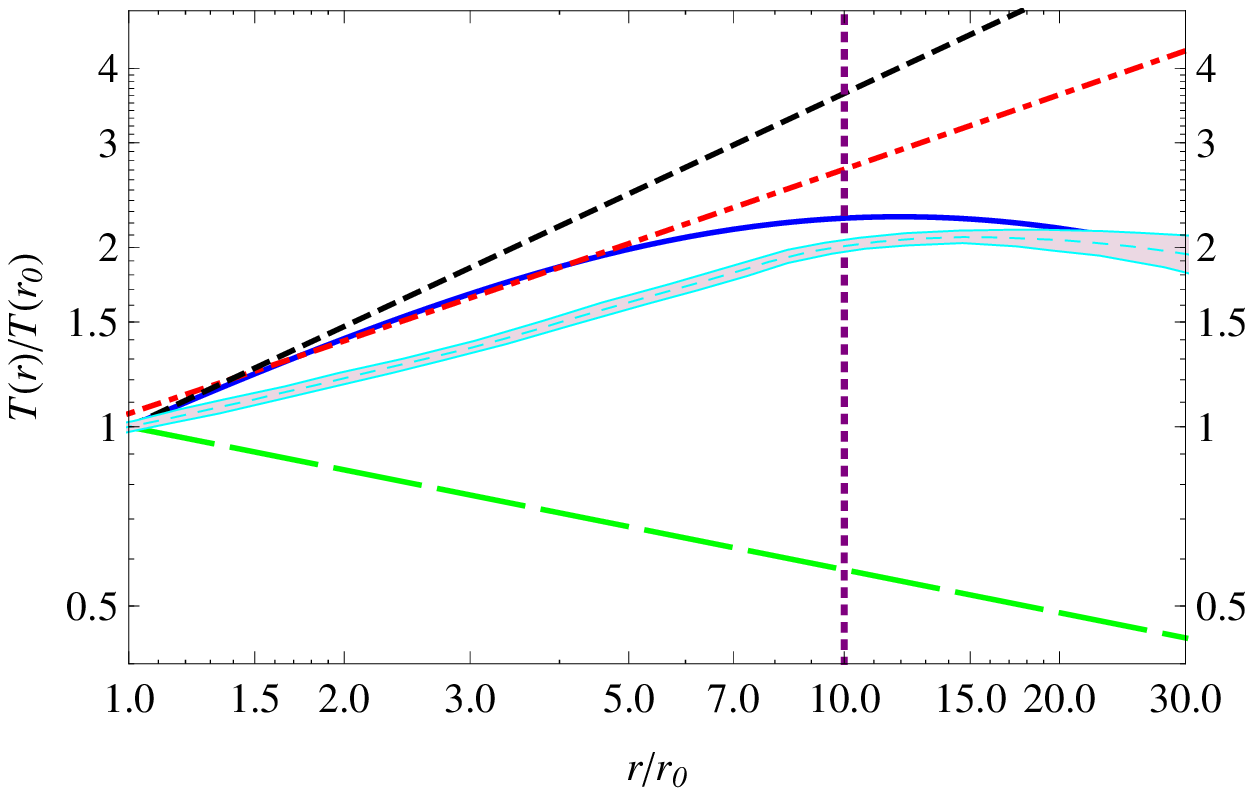}}}
\caption{
Equatorial radial profiles of density (left panel) and temperature (right panel) in the accretion-choked spiral flow model described in \S\ref{sec:Example}.
Shown are the azimuthally averaged profiles (solid curves), their best-fit power-laws $\rho\propto r^{-1.1}$ and $T\propto r^{0.4}$ (dot-dashed curves) evaluated near $r_0$, and the profiles of the dominant, slow (short dashed) and subdominant, fast (long dashed) components.
The model parameters used here are $n_{0.03}r_{10}/T_4=1$ and $\gamma(r_0)=0.2$, or equivalently $\nu=0.05$ and $r_{\mymax}/r_0=10$; see Eqs.~(\ref{eq:r_max}) and (\ref{eq:nu}).
The power-law solution is probably modified beyond $\sim r_{\mymax}$ (dotted horizontal line), unless cooling of the fast component is balanced, for example by heat conduction.
We assume that $\Gamma=5/3$ and $\lambda=-0.6$; in reality the pressure gradient $\mypP$ steepens above $\sim r_{\mymax}$.
The mean deprojected CCC temperature profile found by \citet{SandersonEtAl06} is shown for $r_0=0.01r_{500}$ (dotted curve in the right panel, with a band representing the $1\sigma$ confidence level).
\label{fig:InflowOutflow}
}
\end{figure*}

The azimuthally averaged density profile is given  as a function of the normalized radius $\sigma \equiv r/r_0$ by
\begin{eqnarray}
\rho(\sigma) & \propto & \Delta \phi_f \sigma^{\myprho^{(f)}} + \Delta \phi_s \sigma^{\myprho^{(s)}} \\
& \propto & \left[ 1 - \nu \left( 1-\sigma^{4/5} \right) \right] \sigma^{-\frac{4-3\mypP}{5}} \nonumber \\
& \simeq & \left[ 1 - 0.05 \nu_{0.05} \left( 1-\sigma^{4/5} \right) \right] \sigma^{-1.16} \nonumber \coma
\end{eqnarray}
where $\nu_{0.05}\equiv (\nu/0.05)$, and we used a pressure profile $\mypP\simeq -0.6$ in the last line.
This profile is best fit for the above typical parameters by $\rho\propto \sigma^{-1.1}$ in the range $1<\sigma<10$, as shown in Figure \ref{fig:InflowOutflow} (left panel).

The emissivity-weighted temperature profile is similarly given by
\begin{eqnarray}
T(\sigma) & \propto &
\frac{\Delta \phi_f \sigma^{\frac{3}{2}\mypT^{(f)}+2\myprho^{(f)}} + \Delta \phi_s \sigma^{\frac{3}{2}\mypT^{(s)}+2\myprho^{(s)}}}
{\Delta \phi_f \sigma^{\frac{1}{2}\mypT^{(f)}+2\myprho^{(f)}} + \Delta \phi_s \sigma^{\frac{1}{2}\mypT^{(s)}+2\myprho^{(s)}}} \nonumber \\
& \propto & \frac{1 - \nu \left( 1-\sigma^{2/5} \right)}{1+\nu \left(\sigma^{6/5}-1\right)} \sigma^{\frac{4+2\mypP}{5}}  \\ \nonumber
& \simeq & \frac{1 - 0.05\nu_{0.05} \left( 1-\sigma^{2/5} \right)}{1+0.05\nu_{0.05} \left(\sigma^{6/5}-1\right)} \sigma^{0.56} \fin \nonumber
\end{eqnarray}
The logarithmic slope of $T(\sigma)$ declines with $\sigma$.
The best fit power law for the above typical parameters is $T\propto \sigma^{0.4}$ in the range $1<\sigma<5$, as shown in Figure \ref{fig:InflowOutflow} (right panel).
The figure also shows, for comparison, the universal mean deprojected temperature profile found by \citet{SandersonEtAl06}, scaled with $r_0 = 0.01r_{500}$, where
$r_{500}$ is the radius enclosing 500 times the critical density of the Universe.

These density and temperature profiles are in good agreement with the characteristic structure of cores.
In particular, the temperature peak typically observed near $r_{\mymax}$ is reproduced, although the power-law solution may break down above $\sim r_{\mymax}$ due to cooling and the steepening of the pressure profile.
The azimuthally averaged profiles depend on the volume fraction $\nu$ of the fast flow: smaller (larger) values of $\nu$ lead to a steeper (flatter) density decline and temperature rise, approaching (deviating from) the slow phase profiles shown as black short-dashed curves in Figure \ref{fig:InflowOutflow}.
This may explain the scatter seen around the universal thermal profiles among different clusters.

Various properties of the model can be directly read off Eqs.~(\ref{eq:TRatio})--(\ref{eq:SpiralBase}), and (\ref{eq:nu}), in particular within the equatorial plane.
For example, the fast outflow contains a small, $\nu\sim 5\%$ fraction of the mass at the base of the spiral $r_0$, but an increasing, $[1+(\nu^{-1}-1)(r/r_0)^{-4/5}]^{-1}$ fraction at larger radii.
The kinetic energy of this fast outflow is a small, $\sim (\Gamma \nu/3)(u_f/c_s)^2 \simeq 0.03(r/r_0)^{-8/5}$, radially decreasing fraction of the thermal energy.

Eq.~(\ref{eq:nu}), which guarantees a vanishing mass accretion, similarly implies that the spiral flow carries no angular momentum in the rotating frame, because the slow and fast streamlines are parallel.
The angular momentum of the flow region is therefore $L\simeq \omega I$, where the moment of inertia is $I\simeq (\pi/8)r^2 M(<r)$ in the $\rho\propto r^{-1}$ approximation.

As mentioned above, although spiral flows may carry negligible angular momentum, for example in the choked-accretion case with $\omega\to 0$, the flow necessarily transfers angular momentum radially, because \myNi the angular momentum fluxes of the two components always add up, rather than cancel; and \myNii here the angular momentum flux carried by the fast component strongly dominates that carried by the slow component, by a factor $u_f/u_s\sim 20(\nu/0.05)^{-1}(r/r_0)^{-4/5}$.

Thus, angular momentum in the sense of a leading spiral rotation is transferred outwards, within the equatorial plane and perpendicular to it, towards the poles.
Equivalently, angular momentum in the sense of a trailing spiral rotation is transported inward.
This torques the central region in the trailing direction, for example slowing it down if it initially rotates in a leading sense.
This angular momentum inflow sets a characteristic timescale during which the equatorial rotation velocity at the base of the spiral changes by the sound speed,
\begin{equation} \label{eq:tL}
t_L\sim \left( \frac{r c_s}{\myv_f \myw_f \nu}\right)_{r_0} \sim
1 \left( \frac{ r_{10} }{T_4 \gamma_{0.2} } \right)_{r_0} \nu_{0.05}^{-1} \Gyr \fin
\end{equation}

A significant fraction of the angular momentum inflow may be transferred to the uniform rotation of the spiral flow region, thus extending $t_L$, as demonstrated in \S\ref{sec:Feedback} for the case of torqued bubbles.
However, as the uniform rotation must remain slow, the spiral flow must eventually stop or reverse, unless angular momentum is lost to the cluster's periphery or an additional source of angular momentum torques the center in the opposite, leading sense.

\section{Thermal and mechanical feedback}
\label{sec:Feedback}

An energy source that offsets cooling in the core, and a feedback loop that regulates it, are thought to be essential in quenching the global thermal instability.
AGN output and heat conduction can in principle stem cooling, provided that the feedback efficiency is sufficiently high \citep[\eg][]{GuoEtAl08}, and that the energy is properly distributed in space and time.
Here we discuss the nature of feedback in general terms; a quantitative analysis is beyond the scope of the present study.

Heat conduction and convection can be significantly elevated in the spiral flow region, due to the influence of the bulk flow on the transport processes and the close proximity between hot and cold phases.
Magnetic shear amplification leads to strong magnetic fields parallel to the streamlines, along the TD and within the fast shear layer beneath it, thus suppressing transverse transport \citep{KeshetEtAl10, ZuHoneEtAl11}.
However, parallel transport may be enhanced due to the stronger, aligned fields, with less plasma wave inhibition.
More importantly, the RT unstable layer is not expected to undergo coherent shear; it could support substantial radial heat conduction and convection.

In order to close the feedback loop, the AGN output must be regulated by mass accretion through the slow inspiral, as well as by the fast component if it is an inflow (in the composite inflow case discussed in \S\ref{sec:InflowInflow}).
Although the fast flow is denser, it is not sensitive to heating (or cooling) due to its fast, near-sonic velocity, in particular at small radii.
In contrast, the cooling regulated, slow component is susceptible to heating, but the resulting behavior is sensitive to the details of the heat deposition.
Importantly, both fast and slow components can be significantly modified mechanically, for example by the forces exerted by shocks, jets, and in particular buoyant AGN bubbles.

Consider heating first.
Our model remains valid, with catastrophic cooling being quenched, if heating of the slow inflow is both efficient and able to approximately retain the $\mypv^{(s)}\sim -1/2$ scaling derived above.
This may be achieved, for example, if the specific heating rate roughly scales, on average, as $\dot{\epsilon}(r)\sim r^{-2}$, like the $\Lambda\propto \rho^2 T^{1/2}$ cooling profile.
These requirements are less strict than the $\dot{\epsilon}(\vect{r})\propto \rho^2$ scaling sometimes invoked, as they rely on heat conduction and gas mixing for distributing the heat along streamlines and on small scales.

Next, consider a mechanical forcing of the flow.
Slowing the inflow would only have a cooling effect, as it would increase the ratio between the inspiral time and the cooling time.
In the limit where the forcing is strong enough to cause the infall to change sign and become an outflow, but still remain slow and susceptible to cooling, a steady flow would imply unrealistic thermal profiles, for example $\myprho > \lambda_\tau$ (\cf Eqs.~\ref{eq:Continuity1} and \ref{eq:energy1}).

We conclude that mechanical forces can effectively stem cooling in a (quasi) steady-state flow of the type studied here, only by pushing gas outwards at high velocity.
The fast, outgoing flow may then drag some of the inspiraling gas along, causing the remaining inflow to expand (thus slightly cooling adiabatically), and the cooling rate to decline.
The ultimate outcome, even for an initially pure inflow, is a slow inspiral--fast outspiral, composite flow, of the type discussed in \S\ref{sec:OutflowInflow} and \S\ref{sec:Example}, with an outflow-to-inflow volume ratio $\nu$ regulated by the mechanical feedback.

In this scenario, the inflow--outflow spiral composite flow becomes a direct manifestation of cooling and feedback, in the presence of angular momentum.
The slow inflow, which dominates the core, is in essence a cooling flow, whereas the fast outflow removes dense gas from the center of the core before it has a chance to catastrophically cool.
Both components and the flux ratios between them are regulated by feedback, self-consistently enforced by the spiral-constrained flow.
The steady state here corresponds to a closed circulation with quenched accretion, advecting heat inward and angular momentum (in the leading spiral sense) outward.
The small, few percent ratio between the kinetic energy of the outflow and the thermal energy (see \S\ref{sec:Example}) indicates that this mechanism is energetically plausible.

The rise of AGN bubbles is a natural candidate for mediating the necessary mechanical feedback, as the bubbles preferentially deposit their energy at or just below the strongly magnetized, cool layer.
Preliminary evidence for this is seen in the suggested entrainment of bubbles within the spiral pattern; see Figure \ref{fig:PerseusBubbles}.

The presence of the spiral, with its ordered bulk flow, shear, and magnetic layer, constrains the motion of bubbles through the flow, and facilitates the transfer of kinetic bubble energy to the inflowing gas.
In particular, the spiral directs the bubbles, by channelling them along the magnetic layer and furnishing them with angular momentum in the leading spiral sense.
Thus, the gas encountered by the bubbles acquires more momentum along the outspiral direction, and less random motions, than it would obtain in the absence of a spiral.
This is clear in the extreme case where bubbles are sufficiently large to span a substantial fraction of the cross section area of the inflow, such that gas is either forced outwards or compressed below the bubbles.
This would agree with the large bubble size $R_b$ and its tight, linear scaling with distance, $R_b\simeq 0.6 r$, found by \citet[][see figure 1]{DiehlEtAl08}; note that the inflow cross section is indeed similar in size and in radial dependence, $\sim e^{-2\pi\gamma}r/(1+\nu)\sim 0.7r$ (for a logarithmic spiral;  see \S\ref{sec:Example}).
Conversely, confinement by the spiral magnetic structure can explain the observed size and size--distance relation of bubbles.

This confinement, so far ignored in feedback studies, is likely to play a major role in the flow.
The force that the spiral structure exerts on the bubbles, and the compression of gas below the bubbles, could explain the enhanced pressure observed above the TD, and the deduced trailing rotation of the spiral (see \S\ref{sec:FlowNature}).
It may also explain the high angular momentum of the outflow near the base of the spiral: bubbles are pushed in the leading direction by the magnetic layer already near or below $r_0$, and throughout the spiral, while the more massive spiral structure thus acquires equal angular momentum in the trailing rotation sense.
Thus, once the magnetic spiral structure has formed, \eg by a SASI-like instability, the flow can in principle be sustained (well beyond $t_L$ in Eq.~\ref{eq:tL}) by mechanical feedback without an additional source of angular momentum, and in particular without invoking mergers, if the trailing rotation imparted to the spiral structure is transferred to large radii.
Such a scenario, in which there is neither a net angular momentum nor a net angular momentum flux, is beyond the scope of the present analysis (\eg it probably breaks the uniform rotation assumption), and will be examined numerically in future work.

\section{Observational evidence for two components}
\label{sec:Multiphase}

\begin{figure}[h]
\centerline{\epsfxsize=7cm \epsfbox{\myfig{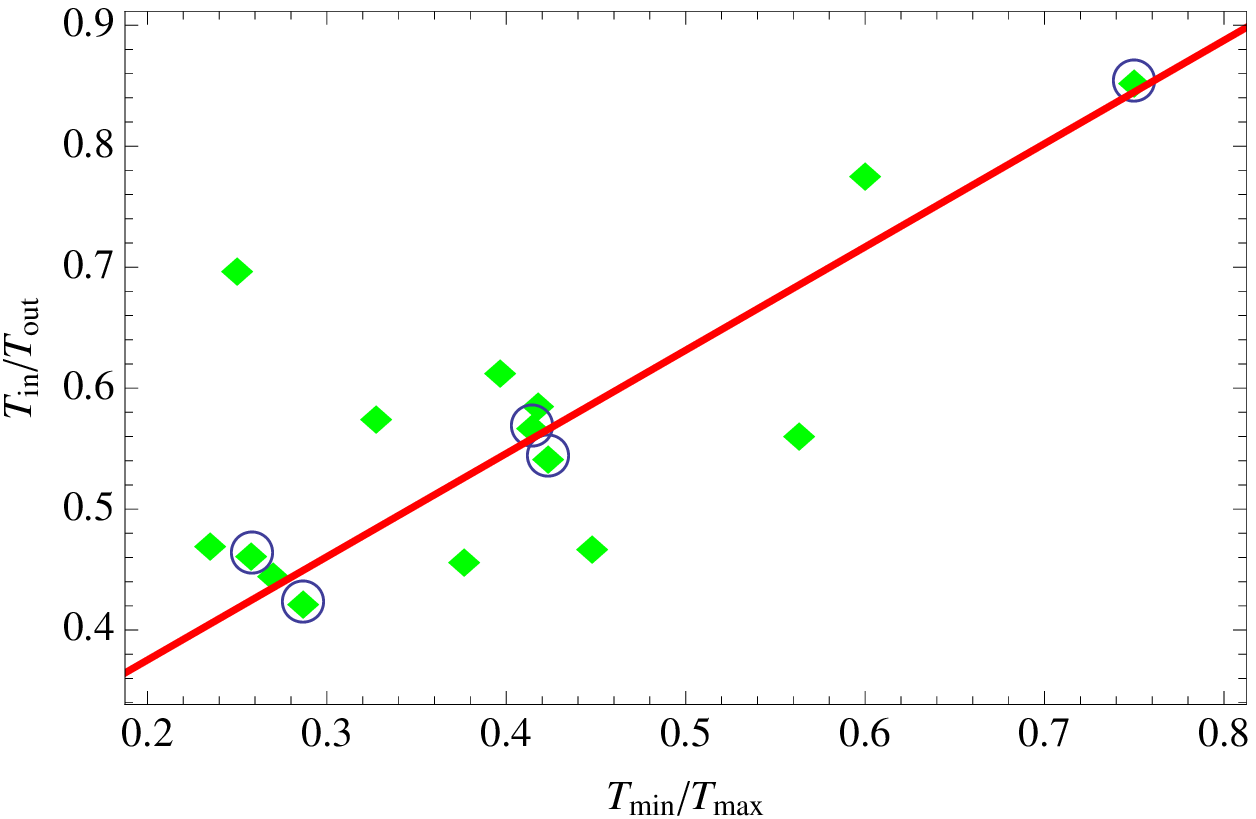}}}
\caption{
Ratio between the inner and outer annulus temperature of the hottest phase in CCs (data from tables 6-7 in \cite{KaastraEtAl04}; Perseus data from \cite{SandersFabian07}), plotted (green diamonds) against the average ratio $T_{\mymin}/T_{\mymax}$ between the coldest and hottest phases in the same annulus (average of best fit column in table 8 of \cite{KaastraEtAl04}).
Higher quality data are highlighted (blue circles with linear fit in red, for clusters with data for $\geq 3$ annuli).
Statistical errors (not shown) are large; we use only data with standard deviation smaller than half the mean.
\label{fig:TRatio}
}
\end{figure}

Observations indicate that CCs typically harbor more than one plasma phase at any given distance from the center of the cluster.
This is seen quite generally, including in clusters where no spiral pattern was so far detected \citep[see, \eg][]{KaastraEtAl04}.
The characteristic temperature spread at a given radius is a factor of $\sim 2-4$ \citep{BuoteEtAl03, KaastraEtAl04}, see Figure \ref{fig:TRatio} (abscissa). An identical spread exists in density, as azimuthal pressure gradients are negligible. A single phase is typically observed only in the outer parts of the core, although this may be due to poor statistics \citep{KaastraEtAl04}.

The azimuthally averaged temperature, and the temperature of the hottest phase, were shown to decline towards the center of the cluster, approximately as a power law, $T\propto r^\mypT$ with $\mypT\simeq 0.4$ \citep{KaastraEtAl04, PiffarettiKaastra06}.

\cite{MolendiPizzolato01} claim that X-ray observations of CCs are consistent at any given radius with a single plasma phase, with azimuthal temperature variations.
There have been attempts to attribute the multiple temperature measurements in some clusters to projection effects through an inherently spherical temperature profile \citep[\eg in A478, ][]{dePlaaEtAl04}.
However, this was shown not to be the case in select cases, for example in 2A 0335+096 \citep{TanakaEtAl06}.
Indeed, in some well-resolved clusters, coherent azimuthal temperature variations of a factor 2--4 are clearly observed, sometimes taking the form of a spiral, as discussed in \S\ref{sec:Intro}.

The emission measure was found to scale roughly as $Y\equiv \int \rho^2\,dV \sim T^\mypY$ with an average $\mypY\sim 4$, in various clusters and different radii \citep{KaastraEtAl04}.
The hotter phase or phases thus strongly dominate the distribution.
For example, for two phases at pressure balance with an average temperature ratio $\langle q\rangle=T_{\mymax}/T_{\mymin}\sim 2$, this would imply a small volume fraction of the cold phase,
\begin{equation}
\nu \sim \left( \frac{T_{\mymin}}{T_{\mymax}} \right)^{\mypY+2} \simeq 0.02 \left( \frac{ \langle q \rangle}{2} \right)^{-6-(\mypY-4)} \coma
\end{equation}
where the result strongly depends upon the poorly constrained $\langle q \rangle$ and $\mypY$.

We point out in Figure \ref{fig:TRatio} that the results of multiphase modelling \citep{KaastraEtAl04} suggest a correlation between the range $T_{\mymin}/T_{\mymax}$ of temperatures observed at a given radius,
and the spatial variation $T_{in}/T_{out}$ of the temperature of the hottest phase across the core.
(However, the statistical errors are substantial.)
Studies of individual clusters also suggest that $T_{\mymin}\sim T_{in}$ \citep[\eg in 2A 0335+096, ][]{TanakaEtAl06}.
Such a correlation is consistent with a model in which the ICM at any radius in the core combines hot plasma from outside the core, and cool plasma of a nearly constant temperature.

\begin{figure}[h]
\centerline{\epsfxsize=7cm \epsfbox{\myfig{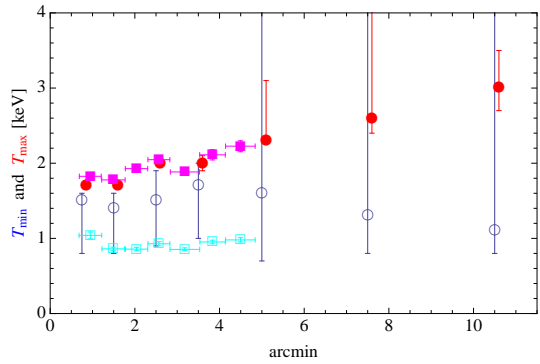}}} \caption{
Projected radial temperature profiles of the hot (filled symbols) and cold (empty symbols) phases in Virgo.
Shown are the azimuthally averaged temperatures (circles; data from tables 6 and 8 of \citet{KaastraEtAl04}), and the temperatures measured along the southwestern X-ray arm (squares; from \citet{WernerEtAl10}).
Here, $1'$ corresponds to $\sim 4.7\kpc$.
\label{fig:TminmaxVirgo}
}
\end{figure}

While the hottest, dominant phase usually becomes colder towards the center of the cluster, the behavior of the cold, subdominant component/s is less clear.
In some clusters, the data are consistent with a cold phase of constant temperature.
This is illustrated for Virgo in Figure \ref{fig:TminmaxVirgo}, using data from \citet[][we chose Virgo because its bins are the most numerous, and amongst the least noisy]{KaastraEtAl04} and \citet{WernerEtAl10}.

The above lines of evidence, combined at face value, suggest azimuthal temperature variations by a factor of 2--4, throughout the core, which is on average strongly dominated by a hot, $T_{\mymax}\sim r^{0.4}$ phase, with a small (of order a few percent by volume) contribution from a cold, $T_{\mymin}\sim\const$ component.
These are indeed the qualitative features derived in the two-component spiral model.
Note, however, that the significance of the evidence reviewed in this section is in general quite low.

\section{Summary and Discussion}
\label{sec:Discussion}

We argue that spiral flows are ubiquitously present in cool cores, and show that the properties of the CCs and of the spiral flows are strongly constrained by the very appearance of piecewise spiral cold fronts in high-resolution X-ray maps.
We interpret such core CFs as projected tangential discontinuities that lie above layers of fast flowing, cold gas, as inferred from the observed CF profiles in \citet{KeshetEtAl10}.

In order to obtain a simple, analytic model, we make several simplifying assumptions/approximations:
\myNi a symmetry axis $z$, with a preferred (equatorial, spiral) perpendicular plane;
\myNii a steady state flow in some rotating (about $\unit{z}$) frame;
\myNiii an ideal gas of adiabatic index $\Gamma=5/3$;
\myNiv a spherically symmetric gravitational potential, \ie in particular, negligible self gravity of the spiral pattern;
\myNv an approximately thermal pressure, nearly (but not exactly) spherically symmetric, scaling as $P\propto r^{\mypP\simeq -0.6}$ (our results depend weakly on $\mypP$);
\myNvi a two-component flow approximation;
\myNvii negligible heat conduction and convection;
\myNviii flow outside the equatorial plane is confined to the plane spanned by the TD normal $\unit{n}$ and $\unit{n}\times\unit{z}$, so the equatorial analysis trivially generalizes to the entire volume;
and
\myNix $\pr_\phi P\sim r^\mypP$.

Under these assumptions and approximations, we find that the TDs and streamlines must follow a bulging cylindrical spiral manifold:
parallel to the equatorial plane, the manifold is approximately a logarithmic (tending slightly towards an Archimedes, $\mypgamma\lesssim 0$) spiral, while perpendicular to the equator it is intermediate between a concentric semicircle and a line parallel to the $z$ axis; see Figures \ref{fig:Defs2} and \ref{fig:Projection}.
Pictorially, one may image such a TD surface as a (rather useless) open barrel, in which the metal hoops were replaced by spiral bands.
The TD parameters (curvature $R_\theta$ perpendicular to the equator, and spiral pattern $\mypgamma$) are summarized in Table \ref{tab:ModelSummary}.
The TD pattern must slowly and uniformly rotate around the $z$-axis.
Subtle pressure gradients (observed at a finite distance above CFs in Perseus) suggest that the spiral is trailing.
The spiral pattern remains apparent in obliquely projected CFs, although it is increasingly distorted and broken to pieces as the viewing axis deviated from $\unit{z}$  (see Figure \ref{fig:Projection}).

We examine the spiral pattern observed in Perseus (see Figures \ref{fig:PerseusPattern1} and \ref{fig:PerseusPattern2}), and the pattern found in the sloshing simulation of \citet[][generated by a major merger with a dark subcluster; see Figures \ref{fig:SPHPattern1} and \ref{fig:SPHPattern2}]{AscasibarMarkevitch06}.
Indeed, in both cases a nearly logarithmic spiral pattern is found.
The pattern in the simulations, in particular, is an approximately uniformly rotating, trailing, logarithmic (slightly tending towards an Archimedes) spiral, in agreement with the model.

In order to account for CF observations, the spiral flow must combine a dense, cold, and metal-rich component, flowing fast and nearly adiabatically in a narrow layer beneath the TD, and a slowly inspiraling, cooling, hotter component, which fills most of the volume.
These two flow components correspond to the two distinct solutions of the angular momentum equation (\ref{eq:momentum_phi1}).
Extrapolating the properties of observed CFs suggest that at some small radius $r_0$, the two components have comparable densities and temperatures, and the fast component is nearly sonic, such that $q\simeq (r/r_0)^{4/5}$ and $u_f/u_s\simeq 20(r/r_0)^{-4/5}$, independent of $\Gamma$ and $\mypP$; see Eqs.~(\ref{eq:TRatio})--(\ref{eq:SpiralBase}) and (\ref{eq:nu}).
A transition layer that separates between the two components at their RT unstable side (where the fast, dense flow lies above the slow flow) mixes the gas and may heat the cold component.

The scaling properties derived for each component of the flow are summarized in Table \ref{tab:ModelSummary}.
The density and temperature of the fast, adiabatic component change slowly with radius, and it is increasingly fast near the center.
In contrast, the velocity of the slow component changes more gradually, and it becomes much denser and somewhat colder at small radii.
The azimuthally averaged density and temperature profiles are dominated by the slow component, and are found to be in good agreement with observations\footnote{One of the two independent thermal parameters is essentially fixed by the gravitational potential; the second is a prediction of the model. Here we choose to externally fix the pressure profile.}: $\rho\sim r^{-1}$ and $T\sim r^{\mypT\lesssim 0.5}$.
The model qualitatively agrees with the multiphase properties observed in CCs (see \S\ref{sec:Multiphase}), and reproduces the typical $\sim 100\kpc$ core size; see Eq.~(\ref{eq:r_max}).

The equations governing the flow do not break a degeneracy in the direction in which the fast, adiabatic component is flowing, allowing for two types of composite spiral flows.
In the first type (see \S\ref{sec:InflowInflow}), both fast and slow components are inflows, such as expected in the sloshing oscillations induced after a merger event disrupts the core (in the presence of cooling).
This is essentially a transient mode, with substantial accretion and little resilience to cooling, unless AGN feedback reverses part of the slow infall and accelerates it outwards to high velocity.
In such a scenario, the core evolves into a second type of flow: an inspiral--outspiral composite mode.

This second type of composite flow (see \S\ref{sec:OutflowInflow}) involves a cold, fast outflow and a hot, slow inflow.
While this requires some mechanism for launching or sustaining the outflow, it naturally solves the global cooling problem with a low level of mass accretion, provided that the outflow is regulated by some accretion-driven feedback mechanism.
AGN bubbles, in particular, provide both a launching mechanism and a plausible feedback loop; see the discussions in \S\ref{sec:Example}, \S\ref{sec:Feedback}, and below.

An additional assumption of choked mass accretion may account for the weak cooling flows and the low star formation rates inferred from observations.
This selects the second, inflow--outflow type of composite spiral flow, strongly constrains the properties of such a flow, and produces thermal profiles in good agreement with observations (see Figure \ref{fig:InflowOutflow}).
Such a flow is essentially a closed circulation, carrying no angular momentum in the rotating frame, advecting heat inward from outside the core, and carrying cold, dense gas outwards at high velocity before it can catastrophically cool.

Both types of spiral flows transfer angular momentum radially, torquing the central core in the sense of a trailing spiral rotation.
This can persist over timescales longer than $t_L\sim 1\Gyr$ (see Eq.~(\ref{eq:tL})) only if the spiral pattern reverses, additional angular momentum is deposited in the center, or the angular momentum involved in the trailing rotation of the entire structure is transferred (\eg magnetically; see \S\ref{sec:Feedback}) outwards.

Both types of flows may quench the local thermal instability, as they efficiently distribute energy spatially and temporally across the core by bringing cold gas from small radii and hot gas from outside the core into close contact, strongly mix the gas in the RT unstable layer, and possibly enhance transport along streamlines.

Moreover, in the presence of AGN feedback, the spiral may stabilize the core also against the global thermal instability, as discussed in \S\ref{sec:Feedback}.
One way this could happen is through heating, but this requires an approximately $r^{-2}$ heating profile.
A more natural and robust feedback involves the mechanical energy output of the AGN, forcing gas outwards at high velocity.
This can lead to stable flows of the second, inflow--outflow type, with realistic core profiles and choked mass accretion.
Here, the spiral plays a central role in coupling the AGN output to the gas, for example by channelling bubbles along the spiral.
This efficiently quenches cooling by regulating the inflow, removing dense gas from the base of the spiral, and modulating the spiral in response to perturbations as to maintain a low level of accretion.
The observed linear, $R_b\sim 0.6r$ scaling of bubble sizes \citep{DiehlEtAl08} may provide direct evidence for such mechanical feedback and coupling.

Regardless of the cooling problem and the type of flow, a spiral flow strongly affects/mediates the AGN output, for example by diverting hot bubbles, distorting jets, and deflecting shocks.
Indeed, there is evidence that bubbles interact with, and appear to be entrained by, the spiral pattern (see for example Figure \ref{fig:PerseusBubbles}; \citet{FormanEtAl07}; \citet{FabianEtAl11});
such rising bubbles do provide direct evidence that at least some part of the ICM is an outflow.
AGN bubbles are therefore natural candidates for mediating the mechanical feedback; this can also explain the trailing spiral rotation.
Interactions with a spiral flow may explain the rapid growth of bubbles and their linear size--distance relation \citep[][and see \S\ref{sec:Example}]{DiehlEtAl08}, the asymmetric deformation of flow around some bubbles \citep{Lim11}, and the large sizes some bubbles attain \citep[\eg][]{McNamaraEtAl05}.

Bubbles and cold filamentary nebulae, observed near the centers of some CCs, gauge the flow.
In particular, coherent axes of bubbles \citep[\eg][]{FabianEtAl11} and extended linear cold filaments, thought to arise from the local thermal instability \citep[\eg][]{PizzolatoSoker06, SharmaEtAl11}, may seem to limit the presence of spiral flows to larger radii.
However,
\myNi most of the volume is filled with the slow, $\lesssim 100\km \se^{-1}$ gas in which differential rotation is minimal, whereas the fast flow is confined to a thin layer below the TD;
\myNii bubbles appear to be entrained by the flow, possibly driving or stabilizing it, so extended, albeit slightly bent, axes are to be expected;
and
\myNiii some extended linear structures may be parallel to the rotation axis $z$, for example the long, warped filaments sometimes observed in isolation.
The presence of filamentary nebulae in some CCs may reflect enhanced cooling due to the weakening of the spiral outflow in those clusters.

By focusing on the steady state flow, we have not addressed the initial formation of the spiral pattern.
However, sloshing simulations have demonstrated that such patterns easily form in cores in response to strong perturbations, and standing accretion shock instability (SASI) simulations suggest that spiral modes are excited even in unperturbed spherical accretion in the presence of cooling.
Note that, if the universal thermal profiles of CCs are indeed the signature of a spiral flow, and if these are composite inflows (of the first type) of a merger origin, then a major merger (of the type invoked to explain the minority, $\sim 1/3$ of non cool-core clusters) would be necessary in order to generate these spirals.
Otherwise, one would expect to see clusters with a disrupted core or no core at all, with no evidence for a major merger.

Groups of galaxies show CCs that resemble those in clusters, with similar profiles and occasionally CFs, but with a higher fraction of gas cooling per unit time \citep[\eg][]{McDonaldEtAl11}.
Our results suggest that such groups too carry spiral flows, but the latter are less efficient in mixing the gas due to the scaling of AGN feedback with system mass.

Our analysis bears important implications for interpreting the structure, composition, and gravitational potential of CCs, showing for example that deviations from hydrostatic equilibrium are locally significant.
Spiral flows may play an important role in mixing the gas on large scales, possibly facilitating the smooth metallicity distribution observed \citep[\eg][]{SandersonEtAl09}, and the homogeneity of cosmic-ray ions inferred from radio observations \citep{Keshet10}.
Our results emphasize that spiral flow studies, such as sloshing simulations, should incorporate AGN feedback, in particular in the form of buoyant bubbles.
Such simulations can directly test if a spiral flow naturally produces the characteristic core structure observed, by studying the evolution and long-term stability of flows with different initial conditions.
Notice that spiral flows cannot be properly simulated without accounting for (or at least mimicking the effect of) the TD-stabilizing magnetic fields.

The existence of spiral flows in cores can be directly tested through the spectroscopic shift and broadening induced by flows along the line of sight.
Evidence for an intricate, possibly spiral velocity structure was thus uncovered in Perseus \citep[][figure 8]{SandersEtAl04}, and evidence for spectral broadening at the level of the sound velocity was found in a small subset of clusters \citep{SandersEtAl11}.
Upper limits on the energy associated with the projected velocity dispersion, at the level of $5\%$ to $20\%$ of the thermal energy, were imposed in several other clusters, by \citet{ChurazovEtAl08, SandersEtAl10, SandersEtAl11, BulbulEtAl12}.
Thus, present spectroscopy is in general unable to probe the narrow, fast outflows, whose kinetic energy is a small, $\sim 0.03(r/r_0)^{-8/5}$ fraction of the thermal energy (see \S\ref{sec:Example}).
In addition, such observations are only sensitive to motions along the line of sight, and typically probe scales larger than the base of the spiral, where the flow is fastest.

The spiral bulk flows should be directly observable with future telescopes, for example using the next international X-ray satellite, ASTRO-H \citep{TakahashiEtAl10}, scheduled for launch in 2014.
In particular, ASTRO-H should be able to spectroscopically resolve the fast flows near the centers of nearby, well-resolved clusters, in which the spiral is not observed face-on (along $\unit{z}$).

In conclusion, we show that the universal thermal profiles of CCs, their size, multiphase properties, common spiral thermal and chemical features, and ubiquitous CFs, can all be explained by the presence of bulk spiral flows.
This suggests that spiral flows may be the fundamental, defining property of CCs.
Such flows can globally stabilize the core against cooling, while quenching mass accretion, by combining a slow, volume-dominant inspiral, with a fast, cold, feedback-regulated outflow.
Present observations are just sensitive to the "tip of the iceberg" of this prevalent, large scale phenomenon.

\acknowledgements
It is a pleasure to thank Y. Birnboim, A. Loeb, N. Soker, W. Forman, P. Sharma, E. Quataert, C.K. Chan, and B. Katz, for encouragement and useful discussions.
Special thanks to M. Markevitch for numerous insightful discussions, and for identifying the X-ray cavities.
This work is supported by NASA through Einstein Postdoctoral Fellowship grant number PF8-90059 awarded by the Chandra X-ray Center, which is operated by the Smithsonian Astrophysical Observatory for NASA under contract NAS8-03060, by a Harvard Institute for Theory and Computation (ITC) fellowship, and by a Marie Curie Reintegration Grant (CIG).

\end{document}

%% file: Definitions.tex
\newcommand{\ie}{\emph{i.e.} }
\newcommand{\eg}{\emph{e.g.,} }
\newcommand{\cf}{\emph{cf.} }

\newcommand{\be}{\begin{equation}}
\newcommand{\ee}{\end{equation}}
\newcommand{\bea}{\begin{equation*}}
\newcommand{\eea}{\end{equation*}}
\newcommand{\beqr}{\begin{eqnarray} \nonumber}
\newcommand{\eeqr}{\end{eqnarray}}
\newcommand{\beqrb}{\begin{eqnarray}}
\newcommand{\eeqrb}{\nonumber \end{eqnarray}}
\newcommand{\fin}{\mbox{ .}}
\newcommand{\coma}{\mbox{ ,}}

\newcommand{\cm}{\mbox{ cm}}

\newcommand{\se}{\mbox{ s}}

\newcommand{\Gyr}{\mbox{ Gyr}}
\newcommand{\erg}{\mbox{ erg}}

\newcommand{\km}{\mbox{ km}}

\newcommand{\kpc}{\mbox{ kpc}}

\newcommand{\keV}{\mbox{ keV}}

\newcommand{\const}{\mbox{const.}}
\newcommand{\constant}{\mbox{constant}}

\newcommand{\vdot}{\mbox{\boldmath{$\cdot$}}}
\newcommand{\X}{\times}

\newcommand{\crl}{\mathbf{\nabla}\X}
\newcommand{\dvr}{\mathbf{\nabla} \vdot}
\newcommand{\grad}{\bm{\nabla}}
\newcommand{\vect}[1]{\mathbf{#1}}
\newcommand{\unit}[1]{\bm{\hat{#1}}}
\newcommand{\pr}{\partial}

\newcommand{\sign}{\mbox{sign}}

\newcommand{\lrgspc}{\,\,\,\,\,\,\,\,\,}
\newcommand{\smlspc}{\,\,\,\,}

\newcommand{\myNi}{\emph{(i)}\,}
\newcommand{\myNii}{\emph{(ii)}\,}
\newcommand{\myNiii}{\emph{(iii)}\,}
\newcommand{\myNiv}{\emph{(iv)}\,}
\newcommand{\myNv}{\emph{(v)}\,}
\newcommand{\myNvi}{\emph{(vi)}\,}
\newcommand{\myNvii}{\emph{(vii)}\,}
\newcommand{\myNviii}{\emph{(viii)}\,}
\newcommand{\myNix}{\emph{(ix)}\,}











%% file: ms.bbl
\begin{thebibliography}{67}
\expandafter\ifx\csname natexlab\endcsname\relax\def\natexlab#1{#1}\fi

\bibitem[{{Ascasibar} \& {Markevitch}(2006)}]{AscasibarMarkevitch06}
{Ascasibar}, Y., \& {Markevitch}, M. 2006, \apj, 650, 102,
  arXiv:astro-ph/0603246

\bibitem[{{Birnboim} {et~al.}(2010){Birnboim}, {Keshet}, \&
  {Hernquist}}]{BirnboimEtAl10}
{Birnboim}, Y., {Keshet}, U., \& {Hernquist}, L. 2010, \mnras, 408, 199,
  1006.1892

\bibitem[{{B{\^i}rzan} {et~al.}(2004){B{\^i}rzan}, {Rafferty}, {McNamara},
  {Wise}, \& {Nulsen}}]{BirzanEtAl04}
{B{\^i}rzan}, L., {Rafferty}, D.~A., {McNamara}, B.~R., {Wise}, M.~W., \&
  {Nulsen}, P.~E.~J. 2004, \apj, 607, 800, arXiv:astro-ph/0402348

\bibitem[{{Blanton} {et~al.}(2011){Blanton}, {Randall}, {Clarke}, {Sarazin},
  {McNamara}, {Douglass}, \& {McDonald}}]{BlantonEtAl11}
{Blanton}, E.~L., {Randall}, S.~W., {Clarke}, T.~E., {Sarazin}, C.~L.,
  {McNamara}, B.~R., {Douglass}, E.~M., \& {McDonald}, M. 2011, \apj, 737, 99,
  1105.4572

\bibitem[{{Blondin} \& {Shaw}(2007)}]{BlondinShaw07}
{Blondin}, J.~M., \& {Shaw}, S. 2007, \apj, 656, 366, arXiv:astro-ph/0611698

\bibitem[{{Bulbul} {et~al.}(2012){Bulbul}, {Smith}, {Foster}, {Cottam},
  {Loewenstein}, {Mushotzky}, \& {Shafer}}]{BulbulEtAl12}
{Bulbul}, G.~E., {Smith}, R.~K., {Foster}, A., {Cottam}, J., {Loewenstein}, M.,
  {Mushotzky}, R., \& {Shafer}, R. 2012, \apj, 747, 32, 1110.4422

\bibitem[{{Buote} {et~al.}(2003){Buote}, {Lewis}, {Brighenti}, \&
  {Mathews}}]{BuoteEtAl03}
{Buote}, D.~A., {Lewis}, A.~D., {Brighenti}, F., \& {Mathews}, W.~G. 2003,
  \apj, 595, 151, arXiv:astro-ph/0303054

\bibitem[{{Churazov} {et~al.}(2003){Churazov}, {Forman}, {Jones}, \&
  {B{\"o}hringer}}]{ChurazovEtAl03}
{Churazov}, E., {Forman}, W., {Jones}, C., \& {B{\"o}hringer}, H. 2003, \apj,
  590, 225, arXiv:astro-ph/0301482

\bibitem[{{Churazov} {et~al.}(2008){Churazov}, {Forman}, {Vikhlinin},
  {Tremaine}, {Gerhard}, \& {Jones}}]{ChurazovEtAl08}
{Churazov}, E., {Forman}, W., {Vikhlinin}, A., {Tremaine}, S., {Gerhard}, O.,
  \& {Jones}, C. 2008, \mnras, 388, 1062, 0711.4686

\bibitem[{{Churazov} {et~al.}(2002){Churazov}, {Sunyaev}, {Forman}, \&
  {B{\"o}hringer}}]{ChurazovEtAl02}
{Churazov}, E., {Sunyaev}, R., {Forman}, W., \& {B{\"o}hringer}, H. 2002,
  \mnras, 332, 729, arXiv:astro-ph/0201125

\bibitem[{{Clarke}(2004)}]{Clarke04}
{Clarke}, T.~E. 2004, Journal of Korean Astronomical Society, 37, 337,
  arXiv:astro-ph/0412268

\bibitem[{{Clarke} {et~al.}(2004){Clarke}, {Blanton}, \&
  {Sarazin}}]{ClarkeEtAl04}
{Clarke}, T.~E., {Blanton}, E.~L., \& {Sarazin}, C.~L. 2004, \apj, 616, 178,
  arXiv:astro-ph/0408068

\bibitem[{{de Plaa} {et~al.}(2004){de Plaa}, {Kaastra}, {Tamura},
  {Pointecouteau}, {Mendez}, \& {Peterson}}]{dePlaaEtAl04}
{de Plaa}, J., {Kaastra}, J.~S., {Tamura}, T., {Pointecouteau}, E., {Mendez},
  M., \& {Peterson}, J.~R. 2004, \aap, 423, 49, arXiv:astro-ph/0405307

\bibitem[{{Diehl} {et~al.}(2008){Diehl}, {Li}, {Fryer}, \&
  {Rafferty}}]{DiehlEtAl08}
{Diehl}, S., {Li}, H., {Fryer}, C.~L., \& {Rafferty}, D. 2008, \apj, 687, 173,
  0801.1825

\bibitem[{{Donahue} {et~al.}(2006){Donahue}, {Horner}, {Cavagnolo}, \&
  {Voit}}]{DonahueEtAl06}
{Donahue}, M., {Horner}, D.~J., {Cavagnolo}, K.~W., \& {Voit}, G.~M. 2006,
  \apj, 643, 730, arXiv:astro-ph/0511401

\bibitem[{{Dunn} \& {Fabian}(2006)}]{DunnFabian06}
{Dunn}, R.~J.~H., \& {Fabian}, A.~C. 2006, \mnras, 373, 959,
  arXiv:astro-ph/0609537

\bibitem[{{Fabian} {et~al.}(2011){Fabian}, {Sanders}, {Allen}, {Canning},
  {Churazov}, {Crawford}, {Forman}, {GaBany}, {Hlavacek-Larrondo}, {Johnstone},
  {Russell}, {Reynolds}, {Salome}, {Taylor}, \& {Young}}]{FabianEtAl11}
{Fabian}, A.~C. {et~al.} 2011, ArXiv e-prints, 1105.5025

\bibitem[{{Fabian} {et~al.}(2006){Fabian}, {Sanders}, {Taylor}, {Allen},
  {Crawford}, {Johnstone}, \& {Iwasawa}}]{FabianEtAl06}
{Fabian}, A.~C., {Sanders}, J.~S., {Taylor}, G.~B., {Allen}, S.~W., {Crawford},
  C.~S., {Johnstone}, R.~M., \& {Iwasawa}, K. 2006, \mnras, 366, 417,
  arXiv:astro-ph/0510476

\bibitem[{{Forman} {et~al.}(2007){Forman}, {Jones}, {Churazov}, {Markevitch},
  {Nulsen}, {Vikhlinin}, {Begelman}, {B{\"o}hringer}, {Eilek}, {Heinz},
  {Kraft}, {Owen}, \& {Pahre}}]{FormanEtAl07}
{Forman}, W. {et~al.} 2007, \apj, 665, 1057, arXiv:astro-ph/0604583

\bibitem[{{Fujita} {et~al.}(2004){Fujita}, {Matsumoto}, \&
  {Wada}}]{FujitaEtAl04}
{Fujita}, Y., {Matsumoto}, T., \& {Wada}, K. 2004, \apjl, 612, L9,
  arXiv:astro-ph/0407368

\bibitem[{{Ghizzardi} {et~al.}(2010){Ghizzardi}, {Rossetti}, \&
  {Molendi}}]{GhizzardiEtAl10}
{Ghizzardi}, S., {Rossetti}, M., \& {Molendi}, S. 2010, \aap, 516, A32+,
  1003.1051

\bibitem[{{Guo} {et~al.}(2008){Guo}, {Oh}, \& {Ruszkowski}}]{GuoEtAl08}
{Guo}, F., {Oh}, S.~P., \& {Ruszkowski}, M. 2008, \apj, 688, 859, 0804.3823

\bibitem[{{Johnson} {et~al.}(2010){Johnson}, {Markevitch}, {Wegner}, {Jones},
  \& {Forman}}]{JohnsonEtAl10}
{Johnson}, R.~E., {Markevitch}, M., {Wegner}, G.~A., {Jones}, C., \& {Forman},
  W.~R. 2010, \apj, 710, 1776, 1001.2441

\bibitem[{{Kaastra} {et~al.}(2004){Kaastra}, {Tamura}, {Peterson}, {Bleeker},
  {Ferrigno}, {Kahn}, {Paerels}, {Piffaretti}, {Branduardi-Raymont}, \&
  {B{\"o}hringer}}]{KaastraEtAl04}
{Kaastra}, J.~S. {et~al.} 2004, \aap, 413, 415, arXiv:astro-ph/0309763

\bibitem[{{Keshet}(2010)}]{Keshet10}
{Keshet}, U. 2010, ArXiv e-prints, 1011.0729

\bibitem[{{Keshet} \& {Loeb}(2010)}]{KeshetLoeb10}
{Keshet}, U., \& {Loeb}, A. 2010, \apj, 722, 737, 1003.1133

\bibitem[{{Keshet} {et~al.}(2010){Keshet}, {Markevitch}, {Birnboim}, \&
  {Loeb}}]{KeshetEtAl10}
{Keshet}, U., {Markevitch}, M., {Birnboim}, Y., \& {Loeb}, A. 2010, \apjl, 719,
  L74

\bibitem[{{Lagan{\'a}} {et~al.}(2010){Lagan{\'a}}, {Andrade-Santos}, \& {Lima
  Neto}}]{LaganaEtAl10}
{Lagan{\'a}}, T.~F., {Andrade-Santos}, F., \& {Lima Neto}, G.~B. 2010, \aap,
  511, A15+, 0911.3785

\bibitem[{{Lim}(2011)}]{Lim11}
{Lim}, J. 2011, in Structure in Clusters and Groups of Galaxies in the Chandra
  Era, 2011 Chandra Science Workshop held July 12-14, 2011. Edited by Jan
  Vrtilek and Paul J. Green. Hosted by the Chandra X-ray Center, Published
  online at: http://cxc.harvard.edu/cdo/xclust11, p.29, ed. {J.~Vrtilek \&
  P.~J.~Green}, 29--+

\bibitem[{{Markevitch} \& {Vikhlinin}(2007)}]{MarkevitchVikhlinin07}
{Markevitch}, M., \& {Vikhlinin}, A. 2007, \physrep, 443, 1,
  arXiv:astro-ph/0701821

\bibitem[{{Markevitch} {et~al.}(2003){Markevitch}, {Vikhlinin}, \&
  {Forman}}]{MarkevitchEtAl03Proc}
{Markevitch}, M., {Vikhlinin}, A., \& {Forman}, W.~R. 2003, in Astronomical
  Society of the Pacific Conference Series, Vol. 301, Astronomical Society of
  the Pacific Conference Series, ed. {S.~Bowyer \& C.-Y.~Hwang}, 37--+

\bibitem[{{Markevitch} {et~al.}(2001){Markevitch}, {Vikhlinin}, \&
  {Mazzotta}}]{MarkevitchEtAl01}
{Markevitch}, M., {Vikhlinin}, A., \& {Mazzotta}, P. 2001, \apjl, 562, L153,
  arXiv:astro-ph/0108520

\bibitem[{{Mazzotta} \& {Giacintucci}(2008)}]{MazzottaGiacintucci08}
{Mazzotta}, P., \& {Giacintucci}, S. 2008, \apjl, 675, L9, 0801.1905

\bibitem[{{Mazzotta} {et~al.}(2001){Mazzotta}, {Markevitch}, {Vikhlinin},
  {Forman}, {David}, \& {VanSpeybroeck}}]{MazzottaEtAl01}
{Mazzotta}, P., {Markevitch}, M., {Vikhlinin}, A., {Forman}, W.~R., {David},
  L.~P., \& {VanSpeybroeck}, L. 2001, \apj, 555, 205, arXiv:astro-ph/0102291

\bibitem[{{McDonald} {et~al.}(2011){McDonald}, {Veilleux}, \&
  {Mushotzky}}]{McDonaldEtAl11}
{McDonald}, M., {Veilleux}, S., \& {Mushotzky}, R. 2011, \apj, 731, 33,
  1102.1972

\bibitem[{{McNamara} \& {Nulsen}(2007)}]{McNamaraNulsen07}
{McNamara}, B.~R., \& {Nulsen}, P.~E.~J. 2007, \araa, 45, 117, 0709.2152

\bibitem[{{McNamara} {et~al.}(2005){McNamara}, {Nulsen}, {Wise}, {Rafferty},
  {Carilli}, {Sarazin}, \& {Blanton}}]{McNamaraEtAl05}
{McNamara}, B.~R., {Nulsen}, P.~E.~J., {Wise}, M.~W., {Rafferty}, D.~A.,
  {Carilli}, C., {Sarazin}, C.~L., \& {Blanton}, E.~L. 2005, \nat, 433, 45

\bibitem[{{Molendi} \& {Pizzolato}(2001)}]{MolendiPizzolato01}
{Molendi}, S., \& {Pizzolato}, F. 2001, \apj, 560, 194, arXiv:astro-ph/0106552

\bibitem[{{Navarro} {et~al.}(1996){Navarro}, {Frenk}, \&
  {White}}]{NavarroEtAl96}
{Navarro}, J.~F., {Frenk}, C.~S., \& {White}, S.~D.~M. 1996, \apj, 462, 563,
  arXiv:astro-ph/9508025

\bibitem[{{O'Dea} {et~al.}(2008){O'Dea}, {Baum}, {Privon}, {Noel-Storr},
  {Quillen}, {Zufelt}, {Park}, {Edge}, {Russell}, {Fabian}, {Donahue},
  {Sarazin}, {McNamara}, {Bregman}, \& {Egami}}]{ODeaEtAl08}
{O'Dea}, C.~P. {et~al.} 2008, \apj, 681, 1035, 0803.1772

\bibitem[{{Peterson} \& {Fabian}(2006)}]{PetersonFabian06}
{Peterson}, J.~R., \& {Fabian}, A.~C. 2006, \physrep, 427, 1,
  arXiv:astro-ph/0512549

\bibitem[{{Piffaretti} \& {Kaastra}(2006)}]{PiffarettiKaastra06}
{Piffaretti}, R., \& {Kaastra}, J.~S. 2006, \aap, 453, 423,
  arXiv:astro-ph/0602376

\bibitem[{{Pistinner} \& {Eichler}(1998)}]{PistinnerEichler98}
{Pistinner}, S.~L., \& {Eichler}, D. 1998, \mnras, 301, 49,
  arXiv:astro-ph/9807025

\bibitem[{{Pizzolato} \& {Soker}(2006)}]{PizzolatoSoker06}
{Pizzolato}, F., \& {Soker}, N. 2006, \mnras, 371, 1835, arXiv:astro-ph/0605534

\bibitem[{{Quataert}(2008)}]{Quataert08}
{Quataert}, E. 2008, \apj, 673, 758, 0710.5521

\bibitem[{{Rafferty} {et~al.}(2008){Rafferty}, {McNamara}, \&
  {Nulsen}}]{RaffertyEtAl08}
{Rafferty}, D.~A., {McNamara}, B.~R., \& {Nulsen}, P.~E.~J. 2008, \apj, 687,
  899, 0802.1864

\bibitem[{{Randall} {et~al.}(2010){Randall}, {Clarke}, {Nulsen}, {Owers},
  {Sarazin}, {Forman}, \& {Murray}}]{RandallEtAl10}
{Randall}, S.~W., {Clarke}, T.~E., {Nulsen}, P.~E.~J., {Owers}, M.~S.,
  {Sarazin}, C.~L., {Forman}, W.~R., \& {Murray}, S.~S. 2010, \apj, 722, 825,
  1008.2921

\bibitem[{{Roediger} {et~al.}(2011){Roediger}, {Br{\"u}ggen}, {Simionescu},
  {B{\"o}hringer}, {Churazov}, \& {Forman}}]{RoedigerEtAl11}
{Roediger}, E., {Br{\"u}ggen}, M., {Simionescu}, A., {B{\"o}hringer}, H.,
  {Churazov}, E., \& {Forman}, W.~R. 2011, \mnras, 413, 2057, 1007.4209

\bibitem[{{Rybicki} \& {Lightman}(1986)}]{RybickiLightman86}
{Rybicki}, G.~B., \& {Lightman}, A.~P. 1986, {Radiative Processes in
  Astrophysics}, ed. {Rybicki, G.~B.~\& Lightman, A.~P.}

\bibitem[{{Sanders} \& {Fabian}(2007)}]{SandersFabian07}
{Sanders}, J.~S., \& {Fabian}, A.~C. 2007, \mnras, 381, 1381, 0705.2712

\bibitem[{{Sanders} {et~al.}(2004){Sanders}, {Fabian}, {Allen}, \&
  {Schmidt}}]{SandersEtAl04}
{Sanders}, J.~S., {Fabian}, A.~C., {Allen}, S.~W., \& {Schmidt}, R.~W. 2004,
  \mnras, 349, 952, arXiv:astro-ph/0311502

\bibitem[{{Sanders} {et~al.}(2011){Sanders}, {Fabian}, \&
  {Smith}}]{SandersEtAl11}
{Sanders}, J.~S., {Fabian}, A.~C., \& {Smith}, R.~K. 2011, \mnras, 410, 1797,
  1008.3500

\bibitem[{{Sanders} {et~al.}(2010){Sanders}, {Fabian}, {Smith}, \&
  {Peterson}}]{SandersEtAl10}
{Sanders}, J.~S., {Fabian}, A.~C., {Smith}, R.~K., \& {Peterson}, J.~R. 2010,
  \mnras, 402, L11, 0911.0763

\bibitem[{{Sanders} {et~al.}(2009){Sanders}, {Fabian}, \&
  {Taylor}}]{SandersEtAl09}
{Sanders}, J.~S., {Fabian}, A.~C., \& {Taylor}, G.~B. 2009, \mnras, 396, 1449,
  0904.1374

\bibitem[{{Sanderson} {et~al.}(2009){Sanderson}, {O'Sullivan}, \&
  {Ponman}}]{SandersonEtAl09}
{Sanderson}, A.~J.~R., {O'Sullivan}, E., \& {Ponman}, T.~J. 2009, \mnras, 395,
  764, 0902.1747

\bibitem[{{Sanderson} {et~al.}(2006){Sanderson}, {Ponman}, \&
  {O'Sullivan}}]{SandersonEtAl06}
{Sanderson}, A.~J.~R., {Ponman}, T.~J., \& {O'Sullivan}, E. 2006, \mnras, 372,
  1496, arXiv:astro-ph/0608423

\bibitem[{{Sharma} {et~al.}(2012){Sharma}, {McCourt}, {Quataert}, \&
  {Parrish}}]{SharmaEtAl11}
{Sharma}, P., {McCourt}, M., {Quataert}, E., \& {Parrish}, I.~J. 2012, \mnras,
  420, 3174, 1106.4816

\bibitem[{{Soker}(2010)}]{Soker10}
{Soker}, N. 2010, ArXiv e-prints, 1007.2249

\bibitem[{{Takahashi} {et~al.}(2010)}]{TakahashiEtAl10}
{Takahashi}, T., {et~al.} 2010, in Society of Photo-Optical Instrumentation
  Engineers (SPIE) Conference Series, Vol. 7732, Society of Photo-Optical
  Instrumentation Engineers (SPIE) Conference Series, 1010.4972

\bibitem[{{Tanaka} {et~al.}(2006){Tanaka}, {Kunieda}, {Hudaverdi}, {Furuzawa},
  \& {Tawara}}]{TanakaEtAl06}
{Tanaka}, T., {Kunieda}, H., {Hudaverdi}, M., {Furuzawa}, A., \& {Tawara}, Y.
  2006, \pasj, 58, 703

\bibitem[{{Tittley} \& {Henriksen}(2005)}]{TittleyHenriksen05}
{Tittley}, E.~R., \& {Henriksen}, M. 2005, \apj, 618, 227,
  arXiv:astro-ph/0409177

\bibitem[{{Vikhlinin} {et~al.}(2006){Vikhlinin}, {Kravtsov}, {Forman}, {Jones},
  {Markevitch}, {Murray}, \& {Van Speybroeck}}]{VikhlininEtAl06}
{Vikhlinin}, A., {Kravtsov}, A., {Forman}, W., {Jones}, C., {Markevitch}, M.,
  {Murray}, S.~S., \& {Van Speybroeck}, L. 2006, \apj, 640, 691,
  arXiv:astro-ph/0507092

\bibitem[{{Voigt} \& {Fabian}(2004)}]{VoigtFabian04}
{Voigt}, L.~M., \& {Fabian}, A.~C. 2004, \mnras, 347, 1130,
  arXiv:astro-ph/0308352

\bibitem[{{Werner} {et~al.}(2010){Werner}, {Simionescu}, {Million}, {Allen},
  {Nulsen}, {von der Linden}, {Hansen}, {B{\"o}hringer}, {Churazov}, {Fabian},
  {Forman}, {Jones}, {Sanders}, \& {Taylor}}]{WernerEtAl10}
{Werner}, N. {et~al.} 2010, \mnras, 407, 2063, 1003.5334

\bibitem[{{Zakamska} \& {Narayan}(2003)}]{ZakamskaNarayan03}
{Zakamska}, N.~L., \& {Narayan}, R. 2003, \apj, 582, 162,
  arXiv:astro-ph/0207127

\bibitem[{{ZuHone} {et~al.}(2010){ZuHone}, {Markevitch}, \&
  {Johnson}}]{ZuHoneEtAl10}
{ZuHone}, J.~A., {Markevitch}, M., \& {Johnson}, R.~E. 2010, \apj, 717, 908,
  0912.0237

\bibitem[{{ZuHone} {et~al.}(2011){ZuHone}, {Markevitch}, \&
  {Lee}}]{ZuHoneEtAl11}
{ZuHone}, J.~A., {Markevitch}, M., \& {Lee}, D. 2011, \apj, 743, 16, 1108.4427

\end{thebibliography}
